\documentclass[amsmath,twocolumn,pra,notitlepage,nofootinbib]{revtex4-2}
\usepackage{amsthm,amsfonts,amssymb,graphicx,verbatim,xfrac,bm}
\usepackage{hyperref}
\usepackage[dvipsnames]{xcolor}
\usepackage[utf8]{inputenc}

\usepackage{tikz} % drawing in-line
\usetikzlibrary{arrows,decorations.pathmorphing,backgrounds,positioning,fit,matrix,calc}
\tikzset{snake it/.style={decorate, decoration=snake,segment length=0.1cm}}

\renewcommand{\vec}[1]{{\bm{#1}}}

\renewcommand{\Re}{{\rm Re}\,}

\newcommand{\ket}[1]{|#1\rangle}

\newcommand{\braOket}[3]{\langle #1|#2|#3\rangle}

\begin{document}
\title{
Unconventional superconductivity due to interband polarization
}
\author{Valentin Cr\'epel$^1$, Tommaso Cea$^2$, Liang Fu$^1$, and Francisco Guinea$^{2,3,4}$}
\affiliation{$^1$ Department of Physics, Massachusetts Institute of Technology, Cambridge, Massachusetts 02139, USA}
\affiliation{$^2$ Imdea Nanoscience, Faraday 9, 28015 Madrid, Spain}
\affiliation{$^3$ Donostia International Physics Center, Paseo Manuel de Lardiz\'abal 4, 20018 San Sebasti\'an, Spain}
\affiliation{$^4$ Ikerbasque. Basque Foundation for Science. 48009 Bilbao. Spain}
 
\begin{abstract}
We analyze in detail the superconductivity that arises in an extended Hubbard model describing a multiband system with repulsive interactions. We show that virtual interband processes induce an effective attractive interaction for small momentum transfers, a situation not found in most models of superconductivity from repulsion. This attraction can be traced back, in real space, to the presence of correlated hopping terms induced by interband polarization. We reveal this physics with both strong-coupling expansion and many-body perturbation theory, supplemented by numerical calculations. Finally, we point out interesting similarities with the problem of interacting electrons in twisted bilayer graphene, suggesting the importance of interband contribution to superconductivity.
\end{abstract}

\maketitle

%\tableofcontents
%\newpage 

\section{Introduction} \label{sec:Introduction}

Superconductivity in low carrier density materials, such as WTe$_2$~\cite{fatemi2018electrically,sajadi2018gate,asaba2018magnetic}, SrTiO$_3$~\cite{schooley1964superconductivity,gastiasoro2020superconductivity}, ZrNCl~\cite{kawaji1997superconductivity,nakagawa2021gate}, twisted bilayer~\cite{cao2018unconventional,lu2019superconductors,yankowitz2019tuning} and trilayer graphene~\cite{chen2019signatures,park2021tunable,zhou2021superconductivity}, is a subject of intense interest. In conventional electron-phonon superconductors, the vast difference between Fermi and Debye energies is responsible for the retarded nature of the phonon-mediated attraction. This allows mobile electrons to interact without being close to each other at the same time, effectively reducing their mutual Coulomb repulsion~\cite{bogoljubov1958new}. In contrast, low density systems lack this large separation of electron and phonon energies, which motivates the study of superconductivity due to the electron-electron repulsion itself, instead of the electron-phonon interaction.

A well-known electronic mechanism for superconductivity has been introduced by Kohn and Luttinger (KL)~\cite{kohn1965new}, in which an effective attraction arises from dynamical screening of the bare repulsion by the highly polarizable electronic Fermi sea~\cite{maiti2013superconductivity,kagan2014kohn}. The KL theory is analytically controlled in the weak interaction regime, and yields a superconducting transition temperature $T_c$ that is orders of magnitude smaller than Fermi temperature $T_F$. In contrast, the experimentally determined $T_c/T_F$ in WTe$_2$, ZrNCl and magic-angle graphene is remarkably high and reaches up to $\sim 0.1$, calling for an electronic mechanism of {\it strong-coupling} superconductivity.

Strong-coupling superconductivity can arise from repulsive interactions, as a result of enhanced fluctuations near insulating ordered phases driven by these interactions. This picture has been extensively discussed in connection to the superconducting cuprates~\cite{KKNUZ15}, pnictides~\cite{S11,YHK14,FC17}, and also twisted bilayer graphene and other graphitic systems~\cite{KIVF19,YV29,LKLHKV19,WS20,KBVZ20,STSVA20,KXM20,CSS21,DHZLM21,CWBZ21, HWQM21}. %A related mechanism~\cite{HWQM21} involves an enhancement of inter-sublattice fluctuations, which leads to a net attraction between quasiparticles at the Fermi level. 
Superconductivity can also arise from repulsive interactions through virtual transition of electron pairs from the Fermi surface to distant or incipient bands~\cite{chen2015electron,dong2021activating}.

Recently, a new mechanism for unconventional superconductivity from repulsive interactions in doped insulators has been proposed, where the attraction between two conduction electrons arises from virtual interband transition of a third electron in the filled band~\cite{CrepelFu_SCfromRepulsion,CrepelFu_TripletSC,slagle2020charge}. 
The pairing interaction obtained by integrating out high-energy interband excitations is non-retarded and acts on all conduction electrons rather than a small fraction near the Fermi level. As a result, the gap-$T_c$ ratio $\Delta/(k_B T_c)$ is significantly higher than the standard value for electron-phonon superconductors, which clearly indicates unconventional strong-coupling superconductivity.  
This “three-particle mechanism” for superconductivity was rigorously demonstrated with a hybridization expansion method~\cite{CrepelFu_SCfromRepulsion}, which is {\it nonperturbative} in interaction strength and provides an analytically controlled theory of strong-coupling superconductivity at low carrier density.

In this work, we present a comprehensive study of the emergence of superconductivity from repulsive interactions in a two-band Hubbard model introduced by two of us in Ref.~\cite{CrepelFu_SCfromRepulsion}. By combining a variety of analytical and numerical methods, our study covers a wide range of parameters and electron fillings. We show that, unlike the Kohn-Luttinger mechanism or fluctuation induced superconductivity discussed above, the screened interaction, when projected onto the states near the Fermi surface, becomes {\it attractive} for small momentum transfers. %, without mediation from other excitations. 
As a result, a nodeless order parameter, similar to the the one induced by phonons, is possible. The existence of this attractive interaction comes from interband screening effects of the bare repulsion, and its effect can be understood in terms of a new interaction, correlated hopping~\cite{CrepelFu_SCfromRepulsion} (also referred to as electron assisted hopping), between conduction electrons. Focus is not only given to the physical origin for electron pairing, but also to the methodology allowing to controllably capture this pairing in various parameter regimes. We finally discuss the connections of this type of superconductivity with the phase diagram of twisted bilayer graphene~\cite{Guinea_pnas18,Ceae2107874118,phong_cm21}.

\section{Illustrative model and outline} \label{sec:Modelandmethods}

We consider spinless fermions on the honeycomb lattice with nearest-neighbor (NN) tunneling $t$, sublattice potential difference $\Delta$ and NN repulsion $V$
\begin{equation} \label{eq:OriginalSpinlessModel}
\mathcal{H} = \Delta \sum_{r \in B} n_r - t \sum_{\langle r, r' \rangle} (c_r^\dagger c_{r'} + c_{r'}^\dagger c_r) + V \sum_{\langle r, r' \rangle} n_r n_{r'} ,
\end{equation}
above unit filling, \textit{i.e.} for $n=1+x$ electrons per unit cell on average, with $0 < x \leq 1$. Ref.~\cite{CrepelFu_SCfromRepulsion} rigorously proved that this model hosts a superconducting phase at low doping above unit filling, assuming $t \ll \Delta$. One goal of this work is to demonstrate the emergence of attractive interaction from bare repulsion through interband polarization in a wide range of dopings $x$ and model parameters $t/\Delta, V/\Delta$. Our work not only extends the results of Ref.~\cite{CrepelFu_SCfromRepulsion} (obtained from kinetic energy expansion) beyond the regime $t\ll \Delta$, but also elucidates the origin of electron pairing from a band picture viewpoint.

To achieve this goal, we first provide, for any parameter set $(t,V,\Delta)$, at least one analytically exact result establishing the presence of attractive interactions between conduction electrons at small doping concentrations. When $t\ll\Delta$ or $V\ll\Delta$, we respectively rely on the previously mentioned hybridization expansion or on standard many-body perturbation theory. Away from these perturbative regime, we can still obtain analytical results if we extend Eq.~\ref{eq:OriginalSpinlessModel} to the case where fermions carry an additional flavor index $\sigma = 1, \cdots , N_f$:
\begin{equation}  \label{eq:NfFlavorModel}
\mathcal{H} = \Delta \sum_{r \in B} n_r - t \sum_{\langle r, r' \rangle, \sigma} (c_{r, \sigma}^\dagger c_{r', \sigma} + hc) + \frac{V}{N_f} \sum_{\langle r, r' \rangle} n_r n_{r'} ,
\end{equation}
where $n_r = \sum_\sigma n_{r,\sigma}$ now denotes the total density at site $r$. Note that we have scaled the NN interaction strength to keep the same balance between kinetic and interaction energies at unit filling $n=1$, which now corresponds to $N_f$ fermions per unit cells. When the number of fermionic flavors goes to infinity ($N_f \gg 1$), the random phase approximation (RPA) provides exact results on the pairing of electrons in the conduction band, for any values of the ratios $t/\Delta$ or $V/\Delta$.

\begin{figure}
\centering
\includegraphics[width=0.9\columnwidth]{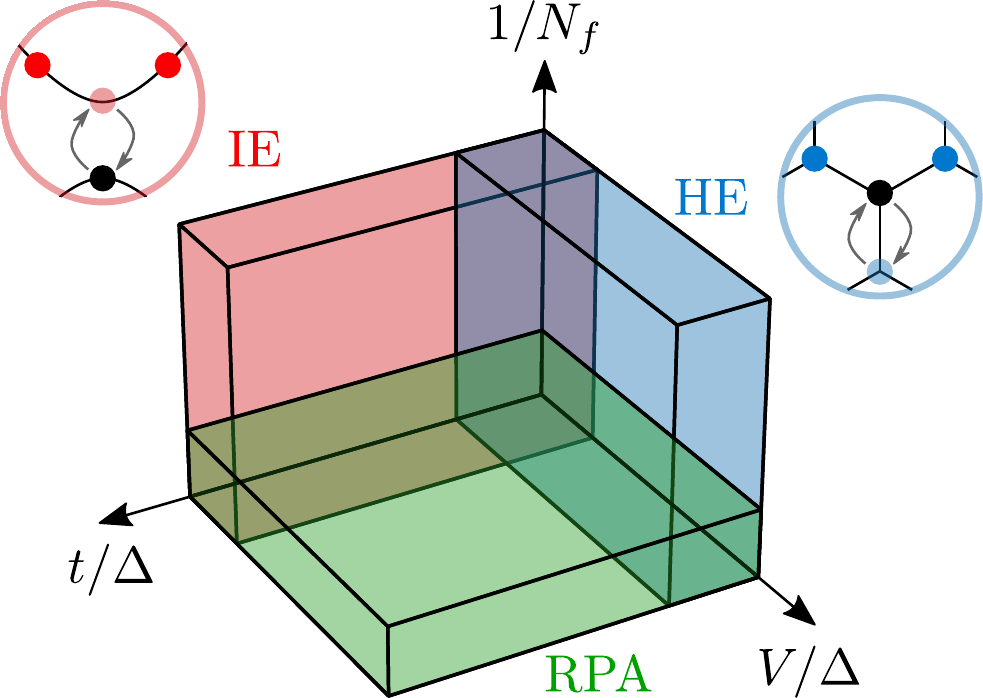}
\caption{Domains of validity of the different analytical methods. The hybridization expansion (HE), interaction expansion (IE) and random phase approximation (RPA) have overlapping regions of validity, allowing for stringent consistency checks. The insets depict the virtual particle-hole processes responsible for pairing, in real-space for the HE ($t\ll\Delta$) and momentum space for the IE ($V\ll \Delta$). 
}
\label{fig:DomainValidity}
\end{figure}

This enterprise is the focus of Secs.~\ref{sec:HybridizationExpansion}-~\ref{sec:InteractionExpansion} and~\ref{sec:RandomPhaseApproximation}, where we derive 
effective models for doped charges and prove the existence of attractive interaction between them due to the interband screening of the bare repulsion. The results obtained in the three regimes of Fig.~\ref{fig:DomainValidity} agree in their overlapping region of validity, which provides a stringent test of our calculations and confirms the emergence of superconductivity in our repulsive model.

We then numerically study the superconducting properties of Eq.~\ref{eq:OriginalSpinlessModel}  as a function of doping in Sec.~\ref{sec:Superconductivity}.  Our simulations include both Hartree-Fock (HF) corrections and dynamical screening effects within the RPA. They show a robust superconducting phase with dome-shaped $T_c$ as a function of doping and a nodeless order parameter for $x<25\%$. Finally, in Sec.~\ref{sec:Discussionconsequences}, we highlight features that transcend our specific model, and discuss their relevance for the superconducting phase of twisted bilayer graphene. Sec.~\ref{sec:conclusion} summarizes and closes the discussion.

\section{Hybridization Expansion} \label{sec:HybridizationExpansion}

We start with the weak tunneling limit $t \ll \Delta$, and briefly summarize the hybridization expansion introduced in Refs.~\cite{CrepelFu_SCfromRepulsion,CrepelFu_TripletSC}, to which we refer for more details.

\subsection{Effective model for doped charges} 

At unit filling and for $t=0$, the system forms a charge density wave in which $A$ sites are occupied with one fermion of each flavor, while the $B$ sites remain empty. States with holes on $A$ sites lie higher in energy due to the large single particle gap $\Delta$. Due to Pauli principle, the $x$ charges added above $n=1$ must live on $B$ sites. For the same reason, direct tunneling of these doped charges to filled adjacent sites is forbidden, and the doped charges' dynamics involves multiple tunneling processes with high energy intermediate state having holes on $A$ sites.

Using a unitary transformation, we can trace over these high-energy excitons to obtain an effective Hamiltonian for the doped charges on the $B$ triangular lattice, thereby including $A-B$ hybridization effects perturbatively in $t/\Delta$. To second order, this hybridization expansion (HE) yields 
\begin{equation} \label{eq:StrongCouplingHamilto_First}
\mathcal{H}^{\rm HE} = \sum_{ijk \in \triangle, \sigma} \left[ c_{i,\sigma}^\dagger T(n_{\triangle}) c_{j,\sigma} + P_{ijk} \right] + E(n_{\triangle})  ,
\end{equation}
where the sum runs over upper triangles $\triangle$ of the $B$ lattice (see App.~\ref{app:HoneycombLattice}). In Eq.~\ref{eq:StrongCouplingHamilto_First}, $n_\triangle = n_i + n_j + n_k$ is the number of fermions in the triangle $\triangle$, and $P_{ijk}$ describes summation over all possible permutations of the indices $ijk$. We have also defined the functions
\begin{subequations} \begin{align}
T(n) & = \frac{t^2}{\Delta + [3 - (2 + n)/N_f] V} , \\
E(n) & = \frac{-t^2 (3N_f - n)}{\Delta + [3 - (1 + n)/N_f] V} ,
\end{align} \end{subequations}
which contain the typical energy difference denominators of second order perturbation theory.

At low density, we can simplify these coefficients assuming that no more than two doped charges simultaneously occupy the same triangle:
\begin{subequations} \begin{align}
T(n) & \simeq T(0) + n [T(1)-T(0)] + \cdots  , \\
E(n) & \simeq E(0) + n [E(1)-E(0)] \\ & \qquad\quad\; + \frac{n(n-1)}{2} [E(2)-2E(1)+E(0)] + \cdots  . \notag
\end{align} \end{subequations}
Defining $t_f = T(0)$, $\lambda = T(1)-T(0)$ and $V_f = E(2)-2E(1)+E(0)$, we can rewrite the effective Hamiltonian for moderate doping as
\begin{equation} \label{eq:StrongCouplingHamilto} \begin{split}
\mathcal{H}^{\rm HE} \simeq & \sum_{ijk \in \triangle, \sigma} \left[ c_{i,\sigma}^\dagger (t_f + \lambda n_\triangle) c_{j,\sigma} + P_{ijk}  \right] \\ 
& + 3 V_f \sum_i \frac{n_i(n_i-1)}{2} + V_f \sum_{\langle i,j\rangle_B} n_i n_j . 
\end{split} \end{equation}
Explicit evaluation of the parameters in this Hamiltonian gives
\begin{subequations} \begin{align}
t_f & = \frac{t^2}{\Delta + [3-2/N_f] V} , \\
\frac{\lambda}{t_f} & = \frac{V / N_f}{\Delta + 3(1-1/N_f) V} , \\
\frac{V_f}{t_f} & = \frac{2 (V/N_f) (\Delta - V/N_f)}{[\Delta + (3-1/N_f) V] [\Delta + 3(1-1/N_f) V]} . 
\end{align} \end{subequations}
They agree with the expression given in  Ref.~\cite{CrepelFu_SCfromRepulsion} and~\cite{CrepelFu_TripletSC} for $N_f=1$ and $N_f=2$, respectively.

\subsection{Dilute limit} 

In the dilute limit, doped electrons mostly occupy states at the band bottom and their physics is governed by long-wavelength properties that transcend details on the lattice scale. To derive this continuum description, we project out the fermionic modes with large kinetic energy and retain only those near the band minima. 

The effective band dispersion of Eq.~\ref{eq:StrongCouplingHamilto} reads $\varepsilon_k = 2t_f \sum_i \cos(k\cdot a_j)$, with $a_j$ the three primitive lattice vectors, and displays two degenerate minima at the $K$ and $K'=-K$ points of the Brillouin zone. At low-energy, doped electrons therefore acquire an additional isospin degree of freedom $\tau$, which distinguishes electrons near $K$ from those close to $K'$, \textit{i.e.} the resulting continuum theory involves $2N_f$ fermionic fields $\psi_{q, \sigma, \tau} = c_{\tau K + q, B, \sigma}$, with $|q a| \ll 1$ and $a$ the lattice constant. Near its minima, $\varepsilon_k$ is quadratic and isotropic, giving the kinetic part of the continuum model 
\begin{equation} \label{eq:ContinuumKineticTerm}
\tilde{\mathcal{H}}_{\rm kin} = \sum_{k,\sigma,\tau} \frac{|k|^2}{2m} \psi_{k, \sigma, \tau}^\dagger \psi_{k, \sigma, \tau} , \;\;  m = \frac{\Delta + (3-2/N_f)V}{3(at)^2} .
\end{equation}

The interaction part of the continuum model is made of the only three symmetry-allowed contact terms -- intra-valley and inter-valley density interactions together with valley-isospin exchange -- which can be written as 
\begin{align} \label{eq:ContinuumInteractionTerm}
\tilde{\mathcal{H}}_{\rm int} & = \frac{1}{N} \sum_{q , \sigma , \sigma', \tau = \pm K }  g_0  \rho_{{q},\sigma,\tau} \rho_{-{q},\sigma',\tau}  \\
& + \frac{1}{N} \sum_{{q}, \sigma, \sigma'} g_1  \rho_{{q},\sigma,K} \rho_{-{q},\sigma',K'} + g_2 \bm{\tau}_{{q}, \sigma} \cdot \bm{\tau}_{-{q},\sigma'} , \notag
\end{align}
where we have defined the density and valley-isospin operators 
\begin{subequations} \begin{align}
\rho_{{q},\sigma,\tau} (q) & = \sum_{{k}} \psi_{{k}+{q}, \sigma ,\tau}^\dagger \psi_{{k}, \sigma, \tau}(q) , \\ 
\bm{\tau}_{{q}, \sigma}(q) & = \sum_{{k},\alpha,\beta} \psi_{{k}+{q}, \sigma, \alpha}^\dagger \left[ \bm{\tau} \right]_{\alpha,\beta} \psi_{{k}, \sigma, \beta} ,
\end{align} \end{subequations}
with $\bm{\tau}$ the set of Pauli matrices. Using the niteraction derived in Eq.~\ref{eq:StrongCouplingHamilto}, we obtain the following $g$-coefficients (see App.~\ref{app:EffectiveContinuum})
\begin{equation}
g_0 =  9 ( V_f - 2 \lambda) /2 , \quad g_1 = 2 g_0 , \quad g_2 = 0  ,
\end{equation}
This drastically simplifies the interacting part of the effective continuum theory, which now simply reads 
\begin{equation}\label{eq:ContinuumContinuumG0only}\tilde{\mathcal{H}} = \sum_{k,\sigma,\tau} \frac{|k|^2}{2m} \psi_{k, \sigma, \tau}^\dagger \psi_{k, \sigma, \tau} + \frac{g_0}{N} \sum_{q} \rho_q \rho_{-q} ,
\end{equation}
with $\rho_q = \sum_{\tau, \sigma} \rho_{q,\sigma,\tau}$ the total density operator. We remark that the original SU($N_f$) flavor symmetry of the model has been promoted to an enlarged SU($2N_f$) symmetry in Eq.~\ref{eq:ContinuumContinuumG0only}, englobing both flavor and the emergent valley degrees of freedom. This is due to the special form of interactions in our model, which does not include any on-site interactions. As shown in Ref.~\cite{CrepelFu_TripletSC}, such on-site repulsion would break the valley isospin rotation symmetry and produce an additional valley-ferromagnetic coupling.

\subsection{Attraction}

Remarkably, $\tilde{\mathcal{H}}$ describes a dilute Fermi gas with \textit{attractive} interactions, as can be seen from the sign of the only relevant interaction coefficient 
\begin{equation} \label{eq:valleysingletenergyKE}
g_0 = - \frac{27 (t V)^2}{N_f P} < 0 , \quad P = \prod_{k=1}^3  [ \Delta + ( 3 - k/N_f ) V ] .
\end{equation}
We stress that this result holds for any values of $N_f$ and is non-perturbative in the interaction parameter $V/\Delta$. Even more importantly, this effective attractive interaction appears for two doped electrons in a band insulator. This strikingly contrast our approach from other mechanisms for superconductivity (such as the KL mechanism) that  rely on the presence of a Fermi surface. It also lead to distinctive predictions, such as a BEC-BCS crossover upon increasing the doping concentration, which has recently been observed in ZrNCl~\cite{nakagawa2021gate, CrepelFu_TripletSC}.

For $N_f=1$, the continuum Hamiltonian for a low density of doped electrons takes the form of a pseudospin-$\frac{1}{2}$ Fermi gas with local {\it attractive} interaction, which is known to exhibit BEC-BCS crossover. On the BCS side and for low densities, the critical temperature satisfies
\begin{equation} \label{eq:TcHybridizationExpansion}
k_B T_c \propto \sqrt{W E_F} e^{-1/(2g)} ,
\end{equation}
where $E_F$ denotes the Fermi energy, $W$ the conduction band width, and $g = 6 V^2 / [\pi \Delta (\Delta + 2V)]$ a dimensionless coupling constant~\cite{CrepelFu_SCfromRepulsion}. Note that $g$ is non-perturbative in the interaction $V$, and it does not have to be small. This allows $k_B T_c/E_F$ ratios as high as 10$\%$. Even for small $g$, since the non-retarded attraction spreads over the entire conduction band, the gap-$T_c$ ratio $\Delta/(k_B T_c)$ is $4.8$, which far exceeds the standard value $1.57$ for electron-phonon superconductors. Both behaviors, obtained by exact solution in the small $t/\Delta$ and low doping regime, clearly indicate strong-coupling superconductivity in our model ~\cite{CrepelFu_SCfromRepulsion}. %, which we asymptotically exact, strong-coupling theory of dilute superconductors that we highlighted in the introduction.

It is also worth comparing Eq.(~\ref{eq:TcHybridizationExpansion}) with the KL result for the two dimensional repulsive Fermi gas~\cite{chubukov1993kohn,raghu2010superconductivity}: $T_c \propto \exp (-1/U_{\rm eff})$, where $U_{\rm eff} \propto U^3$ and $U$ is the strength of bare repulsion. This formula only holds at small $U$, hence  $T_c$ is bound to take values orders of magnitude below the ones of Eq.~\ref{eq:TcHybridizationExpansion}.

The square root dependence of $T_c$ with respect to Fermi energy or doping concentration, shown in Eq.~\ref{eq:TcHybridizationExpansion}, may seem unusual especially given that the density of states near the band edge is constant.  
It originates from the non-retarded attraction between conduction electrons in two dimensions.   
%can be rather simply understood from scattering theory in two dimension within a mean-field treatment akin to BCS~\cite{randeria1990superconductivity}. Indeed, the latter predicts $k_B T_c$ proportional to the zero temperature superconducting gap $\Delta_{T=0}$, which is itself proportional to $\sqrt{E_F \varepsilon_b}$, where $\varepsilon_b$ is the pair binding energy. 
%In 3d, the pair binding energy for a typical interaction strength  $\varepsilon_b \propto E_F e^{-2\pi/(k_F a_s)}$ for a non-retarded attractive potential characterized by its scattering length $a_s$~\cite{randeria1989bound}. As a result, $T_c \propto E_F e^{- \pi/(k_F a_s)}$ is proportional to the Fermi energy, as expected for a fully electronic mechanism in three dimension. In strike contrast, 
In 2D, two-particle bound states exist for arbitrarily weak attraction, and the pair binding energy $\varepsilon_b$ is proportional to the bandwidth, the natural cutoff of the system. At finite density, provided that the Fermi energy $E_F$ is large compared to $\varepsilon_b$ (the BCS regime), both $T_c$ and the zero temperature superconducting gap are proportional to $\sqrt{E_F \varepsilon_b}$, as shown by mean-field treatment detailed in App.~\ref{app:FormulaTc}.

\section{Interaction Expansion} \label{sec:InteractionExpansion}

We now turn to the limit of weak interactions $V \ll \Delta$, for which electrons in the conduction band and holes in the valence band are only mixed slightly. Their small admixture, produced by off-diagonal interactions in the band basis (as opposed to sublattice basis), can be accounted for with standard many-body perturbation theory, as detailed in this section. 

\subsection{Notations}

We rewrite the model Eq.~\ref{eq:NfFlavorModel} in momentum space 
\begin{equation} \label{eq:GenericRGHamiltonianStart} \begin{split}
\mathcal{H} & = \sum_{\substack{12 \\ \sigma}} c_{2,\sigma}^\dagger h_{2,1} c_{1,\sigma} + \frac{1}{2N}  \sum_{\substack{1234  \\ \sigma, \sigma'}} c_{4,\sigma}^\dagger c_{3,\sigma'}^\dagger \Gamma_{43,21} c_{2,\sigma'} c_{1,\sigma} ,
\end{split} \end{equation}
with $N$ the number of unit cells, and $i = ({k}_i, s_i)$ a generalized index gathering the single particle momentum ${k}_i$, belonging to the Brillouin zone (BZ), and the sublattice index $s_i \in \{A,B\}$. Due to momentum conservation, and because our model only features density-density interactions, the parameters of the Hamiltonian can be simplified as 
\begin{equation} \label{eq:SimplerFormInteraction} \begin{split}
h_{2,1} & = \delta_{k_1, k_2} h_{s_2,s_1}(k_1) , \\ \Gamma_{43,21} & = \delta_{s_4,s_1} \delta_{s_3,s_2} \delta_{k_1+k_2,k_3+k_4} \Gamma_{s_2,s_1} (q) ,
\end{split} \end{equation}
with $q = k_4-k_1=k_2-k_3$ the momentum exchanged, and where we have defined the 2-by-2 matrices
\begin{equation}
h ({q}) = \begin{bmatrix} -\Delta/2 & -t f({q}) \\ -tf^*({q}) & \Delta/2
\end{bmatrix} , \; 
\Gamma ({q}) = \frac{V}{N_f} \begin{bmatrix} 0 & f({q}) \\ f^*({q}) & 0
\end{bmatrix} .
\end{equation}
We have used $f({q}) = \sum_{j=1}^3 e^{i({q} \cdot {u}_j)}$, with ${u}_{j=1,2,3}$ the vectors connecting $B$ sites to their three nearest neighbors (see App.~\ref{app:HoneycombLattice}). The one-body part $h$ gives two bands with dispersion
\begin{equation} \label{eq:SingleParticleEigenEnergies}
\varepsilon_{{q}, n} = n \sqrt{(\Delta/2)^2 + |t f({q})|^2} , 
\end{equation}
where $n=\pm$ for the valence and conduction band, respectively. As in Sec.~\ref{sec:HybridizationExpansion}, the conduction band displays degenerate minima at the $K$ and $K'$ points. The corresponding Bloch wavefunctions are
\begin{equation} \label{eq:SingleParticleBloch}
\Psi_{{q},n} = \frac{1}{\sqrt{2\varepsilon_{{q},+} (\varepsilon_{{q},+} + n \Delta/2)}} \begin{bmatrix} - n t f({q}) \\ \varepsilon_{{q},+} + n \Delta/2
\end{bmatrix} ,
\end{equation}
whose upper/lower component corresponds to the wavefunction's amplitude on each sublattices $\Psi_{{q},n}^A$/$\Psi_{{q},n}^B$.

\subsection{Kohn Luttinger diagrams}

Renormalization of the two-body scattering vertex is described, to second order in many-body perturbation theory, by the five diagrams~\cite{kohn1965new}
\begin{equation} \label{eq:FiveKLDiagrams}
\begin{tikzpicture}[baseline=0cm,scale=0.8]
\draw (-0.75,0.5) -- (0.75,0.5); \draw (-0.75,-0.5) -- (0.75,-0.5); 
\draw[snake it] (-0.33,0.5) -- (-0.33,-0.5); \draw[snake it] (0.33,0.5) -- (0.33,-0.5);
\end{tikzpicture} , \, 
\begin{tikzpicture}[baseline=0cm,scale=0.8]
\draw (-0.75,0.5) -- (0.75,0.5); \draw (-0.75,-0.5) -- (0.75,-0.5); 
\draw[snake it] (-0.33,0.5) -- (0.33,-0.5); \draw[snake it] (0.33,0.5) -- (-0.33,-0.5);
\end{tikzpicture} , \, 
\begin{tikzpicture}[baseline=0cm,scale=0.8]
\draw (-0.75,-0.5) -- (0.75,-0.5); \draw (-0.75,0.5) -- (-0.33,0.5) -- (0,0) -- (0.33,0.5) -- (0.75,0.5); 
\draw[snake it] (0,-0.5) -- (0,0); \draw[snake it] (0.33,0.5) -- (-0.33,0.5);
\end{tikzpicture} , \, 
\begin{tikzpicture}[baseline=0cm,scale=0.8]
\draw (-0.75,0.5) -- (0.75,0.5); \draw (-0.75,-0.5) -- (-0.33,-0.5) -- (0,0) -- (0.33,-0.5) -- (0.75,-0.5); 
\draw[snake it] (0,0.5) -- (0,0); \draw[snake it] (0.33,-0.5) -- (-0.33,-0.5);
\end{tikzpicture} , \, 
\begin{tikzpicture}[baseline=0cm,scale=0.8]
\draw (0,0) circle (0.25); 
\draw (-0.75,0.5) -- (0.75,0.5); \draw (-0.75,-0.5) -- (0.75,-0.5); 
\draw[snake it] (0.,0.5) -- (0.,0.25); \draw[snake it] (0.,-0.5) -- (0,-0.25);
\end{tikzpicture} , \, 
\end{equation}
where curvy and straight lines respectively denote interaction events and single-particle propagators. The evaluation of these diagrams is detailed in App.~\ref{app:KLdiagrams}, and produces a renormalized interaction vertex $\Gamma^{\rm IE} = \Gamma + \delta \Gamma$, with
\begin{widetext} \begin{equation} \label{eq:AllKLTerms}
\delta \Gamma_{43,21} = - \frac{1}{N} \sum_{abcd} \chi_{dc,ba}^- \Gamma_{43,ba} \Gamma_{dc,21} + \chi_{dc,ba}^+ [\Gamma_{4c,2a} \Gamma_{d3,b1} + \Gamma_{c3,2a} \Gamma_{d4,b1} + \Gamma_{4c,a1} \Gamma_{d3,b2} - N_f \Gamma_{4c,a1} \Gamma_{d3,2b}] , 
\end{equation} \end{widetext}
where the terms are ordered as in Eq.~\ref{eq:FiveKLDiagrams}. We recall that the sum over the generalized indices $(a,b,c,d)$ runs over all single particle momenta $(k_a,k_b,k_c,k_d )\in {\rm BZ}$ and orbital indices $(s_a,s_b,s_c,s_d) \in \{A,B\}$ that characterize the single particle propagators through the Bloch eigenvectors $\Psi_{k_i,n}^{s_i}$. Finally, we have denoted as $\chi^-$ and $\chi^+$ the particle-particle and particle-hole susceptibilities, respectively. They assume the explicit form
\begin{align}\label{eq:SusceptibilitiesKL}
& \chi_{dc,ba}^{\epsilon = \pm} = \delta_{{k}_a}^{{k}_d} \delta_{{k}_b}^{{k}_c} \\
& \times \sum_{n,n'=\pm} \Psi_{k_d,n}^{s_d \, *} \Psi_{k_c,n'}^{s_c \, *} \Psi_{k_b,n'}^{s_b} \Psi_{k_a,n}^{s_a} \frac{f_\beta (\epsilon \xi_{{k}_a,n}) - f_\beta(\xi_{{k}_b,n'})}{\xi_{{k}_a,n} - \epsilon \xi_{{k}_b,n'}} , \notag
\end{align}
where $f_\beta(x) = 1/[1+e^{\beta x}]$ is the Fermi-Dirac distribution, and $\xi_i = \varepsilon_i-\mu$ measures energies with respect to the chemical potential $\mu$.

The diagrams of Eq.~\ref{eq:FiveKLDiagrams} were first used by Kohn and Luttinger to show how attractive interactions could be produced in a fully repulsive Fermi sea, as a consequence of the dynamical screening of the bare repulsion by the Fermi sea~\cite{kohn1965new}. This picture relies on a large particle-hole susceptibility near $2k_F$, which requires a fully formed Fermi sea. At infinitesimal doping however, this condition is not satisfied and the susceptibility almost vanishes, leading to a breakdown of the KL mechanism. 

Here, in contrast, the particle-hole susceptibility remains finite at low density at low densities, and is dominated by interband contributions. To see this, consider a situation where the chemical potential is positioned right at the conduction band bottom, such that $n=1+x$ with $x\ll1$. At low temperature, the valence band is completely filled $f_\beta(\xi_{{k},-}) = 1$ and the conduction band barely populated $f_\beta(\xi_{{k},+}) \simeq 0$. This forces the band indices in the particle-hole susceptibility to be opposite, thereby highlighting the crucial role of virtual interband transitions.

\subsection{Pairing in the dilute limit}

As in Sec.~\ref{sec:HybridizationExpansion}, we unveil the low density properties of the model with an effective continuum field theory that only retains the modes near the conduction band minima. Following similar lines as above (see App.~\ref{app:EffectiveContinuum}), we end up with the continuum model of Eqs.~\ref{eq:ContinuumKineticTerm} and~\ref{eq:ContinuumInteractionTerm}, albeit with a new mass $m = \Delta / [3(at)^2]$, which agrees with Eq.~\ref{eq:ContinuumKineticTerm} when $V \ll \Delta$, and new $g$-coefficients
\begin{subequations} \label{eq:InteractionExpansionGCoeff} \begin{eqnarray}
g_0 &=& \Lambda_{K,K}(0) - \Lambda_{K,K}(K)/2 , \\ 
g_1 &=& 2\Lambda_{K,K'}(0) + \Lambda_{K,K}(K) , \\ 
g_2 &=& 2 \Lambda_{K,K}(K) ,
\end{eqnarray} \end{subequations}
where $\Lambda$ denotes the effective interaction in the band basis 
\begin{equation}
\Lambda_{k,k'}(q) = \frac{1}{2}  \sum_{a,b = A/B} \!\!\!\!\! \Psi_{{k}+{q},+}^{b \, *} \Psi_{{k}',+}^{a \, *} \Gamma_{a,b}^{\rm IE} ({q}) \Psi_{{k}'+{q},+}^{a} \Psi_{{k},+}^{b} .
\end{equation}

A few simple observations allow us to greatly simplify the expressions of the $g$-coefficients. First, electrons near the conduction band bottom (at $K$ and $K'$) have a strong $B$ character since $\Psi_{\pm K,+}^{A} = 0$. Because all the diagrams in the $g$'s have incoming/outcoming electrons with momentum $\pm K$, this forces all external legs in these diagrams hold a $B$ sublattice index. Furthermore, the bare propagator does not have any direct $B-B$ interaction ($\Gamma_{B,B}(k) = 0$). Together, these observations already rule out the presence of first order interactions at the conduction band bottom, requiring to go to second order and evaluate the diagrams of Eq.~\ref{eq:FiveKLDiagrams}.

Due to the special form of interactions (Eq.~\ref{eq:SimplerFormInteraction}), the orbital index can only change along the single particle propagators (straight lines) of these diagrams. This readily shows that the first four KL diagrams vanish in the dilute limit. To see this, we can draw all diagrams compatible with the above orbital rules, using red/green for A/B indices
\begin{equation}
\begin{tikzpicture}[baseline=0cm,scale=0.8]
\draw[color=OliveGreen,line width=1.25] (-0.75,0.5) -- (0.75,0.5); \draw[color=OliveGreen,line width=1.25] (-0.75,-0.5) -- (0.75,-0.5); 
\draw[snake it] (-0.33,0.5) -- (-0.33,-0.5); \draw[snake it] (0.33,0.5) -- (0.33,-0.5);
\end{tikzpicture} , \, 
\begin{tikzpicture}[baseline=0cm,scale=0.8]
\draw[color=OliveGreen,line width=1.25] (-0.75,0.5) -- (0.75,0.5); \draw[color=OliveGreen,line width=1.25] (-0.75,-0.5) -- (0.75,-0.5); 
\draw[snake it] (-0.33,0.5) -- (0.33,-0.5); \draw[snake it] (0.33,0.5) -- (-0.33,-0.5);
\end{tikzpicture} , \, 
\begin{tikzpicture}[baseline=0cm,scale=0.8]
\draw[color=OliveGreen,line width=1.25] (-0.75,-0.5) -- (0.75,-0.5); \draw[color=OliveGreen,line width=1.25] (-0.75,0.5) -- (-0.33,0.5) -- (0,0) -- (0.33,0.5) -- (0.75,0.5); 
\draw[snake it] (0,-0.5) -- (0,0); \draw[snake it] (0.33,0.5) -- (-0.33,0.5);
\end{tikzpicture} , \, 
\begin{tikzpicture}[baseline=0cm,scale=0.8]
\draw[color=OliveGreen,line width=1.25] (-0.75,-0.5) -- (0.75,-0.5); \draw[color=OliveGreen,line width=1.25] (-0.75,0.5) -- (-0.33,0.5) -- (-0.165,0.25); \draw[color=OliveGreen,line width=1.25] (0.75,0.5) -- (0.33,0.5) -- (0.165,0.25); \draw[color=BrickRed,line width=1.25] (0.165,0.25) -- (0.,0.) -- (-0.165,0.25);
\draw[snake it] (0,-0.5) -- (0,0); \draw[snake it] (0.33,0.5) -- (-0.33,0.5);
\end{tikzpicture} , \, 
\begin{tikzpicture}[baseline=0cm,scale=0.8]
\draw[color=OliveGreen,line width=1.25] (-0.75,0.5) -- (0.75,0.5); \draw[color=OliveGreen,line width=1.25] (-0.75,-0.5) -- (-0.33,-0.5) -- (0,0) -- (0.33,-0.5) -- (0.75,-0.5); 
\draw[snake it] (0,0.5) -- (0,0); \draw[snake it] (0.33,-0.5) -- (-0.33,-0.5);
\end{tikzpicture} , \, 
\begin{tikzpicture}[baseline=0cm,scale=0.8]
\draw[color=OliveGreen,line width=1.25] (-0.75,0.5) -- (0.75,0.5); \draw[color=OliveGreen,line width=1.25] (-0.75,-0.5) -- (-0.33,-0.5) -- (-0.165,-0.25); \draw[color=OliveGreen,line width=1.25] (0.75,-0.5) -- (0.33,-0.5) -- (0.165,-0.25); \draw[color=BrickRed,line width=1.25] (0.165,-0.25) -- (0.,0.) -- (-0.165,-0.25);
\draw[snake it] (0,0.5) -- (0,0); \draw[snake it] (0.33,-0.5) -- (-0.33,-0.5);
\end{tikzpicture} .
\end{equation}
They all feature at least one $B-B$ bare interaction, reducing them to zero. As a result, only the bubble polarization diagram 
\begin{equation} \label{eq:bubbleSpecificpairing}
\begin{tikzpicture}[baseline=0cm,scale=0.8]
\draw[color=BrickRed,line width=1.25] (0,0) circle (0.25); 
\draw[color=OliveGreen,line width=1.25] (-0.75,0.75) -- (0.75,0.75); \draw[color=OliveGreen,line width=1.25] (-0.75,-0.75) -- (0.75,-0.75); 
\draw[snake it] (0.,0.75) -- (0.,0.25); \draw[snake it] (0.,-0.75) -- (0,-0.25);
\node at (-0.6,0.5) {\scriptsize $q=0$}; \node at (-0.6,-0.5) {\scriptsize $q=0$}; \node[color=BrickRed] at (1.1,0) {\scriptsize $\chi^0(q=0)$};
\end{tikzpicture} , \, 
\end{equation}
remains in the weakly interacting and low density limit. Its interaction vertices can, in principle, involve the momentum exchange $q=0$ or $q=\pm K$. However, because the bare interaction vanishes for $\Gamma_{A,B}(\pm K) = V f(\pm K)/N_f = 0$, only the diagram with $q=0$, shown in Eq.~\ref{eq:bubbleSpecificpairing}, is non-vanishing. We extend the discussion on the contributions of the five KL diagrams in the low density limit in App.~\ref{app:diagrams}.

This shows that $\Lambda_{K,K}(K) = 0$, implying $g_2=0$. Because the bubble contribution only depends on the momentum exchanged $q$, we find $\Lambda_{K,K}(0)=\Lambda_{K,K'}(0)$ and therefore $g_1 = 2 g_0$. As a consequence, the effective continuum theory is the same as Eq.~\ref{eq:ContinuumContinuumG0only}. The sign of $g_0 = \Lambda_{K,K}(0) = \Gamma_{B,B}^{\rm IE} (0)/2 = \delta \Gamma_{B,B}^{\rm bbl} (0)/2$ determines the nature of interactions, attractive or repulsive.

To distinguish between these two possibilities, we evaluate the contribution of the bubble diagram in Eq.~\ref{eq:bubbleSpecificpairing}. We recall that the interactions of our original model can be written as $\mathcal{H}_{\rm int} = (2N)^{-1} \sum_{s_1,s_2 \in \{A,B\}} \Gamma_{s_2,s_1}(q) n_{s_1}(q) n_{s_2}(-q)$ with $\Gamma_{s_1,s_2}$ a 2-by-2 matrix. The bubble contribution preserves this matrix structure, yielding
\begin{equation} \label{eq:AuxBblForRPA}
\delta \Gamma_{s_2,s_1}^{\rm bbl} (q) =  N_f \sum_{s_a,s_b} \Gamma_{s_2,s_b} (q) \chi_{s_b,s_a}^0 (q) \Gamma_{s_a,s_1} (q) ,
\end{equation}
with 
\begin{equation} \label{eq:RPAsuscdef}
\chi_{s',s}^0 (q) = \frac{1}{N} \sum_k \chi_{(k,s')(k+q,s),(k+q,s')(k,s)}^+ .
\end{equation}
More succinctly, it can be written as a matrix product $\delta \Gamma^{\rm bbl} (q) = N_f \Gamma (q) \chi^0 (q) \Gamma (q)$. As described above, in the dilute limit, the bubble polarization $\chi^0$ is dominated by interband contributions, which give 
\begin{equation}
\chi \equiv \chi_{A,A}^0 (0) = -\chi_{A,B}^0 (0) = - \frac{1}{N} \sum_{{k}} \frac{|t f({k})|^2}{4 \varepsilon_{{k},+}^3} ,
\end{equation}
and $\chi_{B,B}^0 (0)=\chi_{A,A}^0 (0)$, $\chi_{A,B}^0 (0)=\chi_{A,B}^0 (0)$. 

Putting all pieces together, we find -- as originally announced -- an effective \textit{attractive} interaction in the dilute limit
\begin{equation} \label{eq:valleysingletenergyKL}
g_0 =  - \frac{9 V^2 |\chi|}{2 N_f} . 
\end{equation}
As a consistency check of this result, we can apply it to the case $t\ll\Delta$ where our hybridization expansion result also holds. In that limit, we use the identity $N^{-1} \sum_k |f(k)|^2 = 3$ to get $g_0 \simeq - 27 (tV)^2 / [N_f \Delta^3]$, which perfectly match Eq.~\ref{eq:valleysingletenergyKE} when $V \ll \Delta$ (for which $P \simeq \Delta^3$).

\section{Random phase approximation} \label{sec:RandomPhaseApproximation}

The hybridization and interaction expansions, which are real-space and momentum-space based methods respectively, allow us to solve our extended Hubbard model in different parts of the parameter space. 
We can further bridge between the two approaches and obtain analytical results for any values of $t$, $\Delta$ and $V$ in the limit where $N_f \to \infty$ using the random phase approximation (RPA). 

\subsection{Hartree-Fock and RPA diagrams}

The essence of RPA lies in the possibility to exactly sum the entire series of diagrams that dominate in the $N_f \gg 1$ limit. Such dominating diagrams maximize the number of bubbles for a given number of interaction vertices because, for flavor-conserving interactions, each bubble multiplies its diagram's contribution by $N_f$, while other configurations yield smaller combinatorial factors. We have already encountered this feature in Eq.~\ref{eq:AllKLTerms}, where only the bubble diagram is mutliplied by a factor $N_f$. Generalizing to all orders, the only relevant diagrams in the $N_f \to \infty$ limit are
\begin{equation} \label{eq:RPAseriesofdiagrams}
\begin{tikzpicture}[baseline=0cm,scale=0.8]
\draw (-0.5,0.5) -- (0.5,0.5); \draw (-0.5,-0.5) -- (0.5,-0.5); 
\draw[snake it] (0.,0.5) -- (0.,-0.5);
\end{tikzpicture} , \;
\begin{tikzpicture}[baseline=0cm,scale=0.8]
\draw (0,0) circle (0.25); 
\draw (-0.5,0.5) -- (0.5,0.5); \draw (-0.5,-0.5) -- (0.5,-0.5); 
\draw[snake it] (0.,0.5) -- (0.,0.25); \draw[snake it] (0.,-0.5) -- (0,-0.25);
\end{tikzpicture} , \; 
\begin{tikzpicture}[baseline=0cm,scale=0.8]
\draw (0,0) circle (0.25); \draw (0.75,0) circle (0.25); 
\draw (-0.5,0.5) -- (1.25,0.5); \draw (-0.5,-0.5) -- (1.25,-0.5); 
\draw[snake it] (0.,0.5) -- (0.,0.25); \draw[snake it] (0.75,-0.5) -- (0.75,-0.25); \draw[snake it] (0.25,0) -- (0.5,0); 
\end{tikzpicture} , \;  
\begin{tikzpicture}[baseline=0cm,scale=0.8]
\draw (0,0) circle (0.25); \draw (0.75,0) circle (0.25); \draw (1.5,0) circle (0.25);
\draw (-0.5,0.5) -- (2.,0.5); \draw (-0.5,-0.5) -- (2.,-0.5); 
\draw[snake it] (0.,0.5) -- (0.,0.25); \draw[snake it] (1.5,-0.5) -- (1.5,-0.25); \draw[snake it] (0.25,0) -- (0.5,0); \draw[snake it] (1.,0) -- (1.25,0); 
\end{tikzpicture} , \; \cdots .
\end{equation}

This diagrammatic series can be recast in a geometric form, enabling its exact summation. To see this, recall that the one-bubble correction takes the simple matrix product form $\delta \Gamma^{\rm bbl} (q) = N_f \Gamma (q) \chi^0 (q) \Gamma (q)$. Its obtention can be generalized to the $m$-bubble diagram, giving $(N_f \Gamma \chi^0)^m \Gamma$. Summing all these corrections yields the renormalized RPA scattering vertex 
\begin{equation} \label{eq:cRPA_DefitnitionRPA}
\Gamma^{\rm RPA} (q) = [1-N_f \Gamma (q) \chi^0(q)]^{-1} \Gamma (q) ,
\end{equation}
which is valid when $N_f \gg 1$. Performing the matrix inversion explicitly, we get
\begin{equation} \label{eq:cRPA_ExplicitRPA}
\Gamma^{\rm RPA} = \frac{1}{D} \left( \Gamma + \frac{|V f|^2}{N_f} \begin{bmatrix} \chi_{BB}^0 & -\chi_{AB}^0 \\ -\chi_{BA}^0 & \chi_{AA}^0 \end{bmatrix} \right) ,
\end{equation}
with $D = (1-Vf\chi_{BA}^0)(1-Vf^*\chi_{AB}^0) - |Vf|^2 \chi_{AA}^0 \chi_{BB}^0$. For convenience of notation, the $q$ dependence is not explicitly shown.

\begin{figure*}
\centering
\includegraphics[width=0.7\columnwidth]{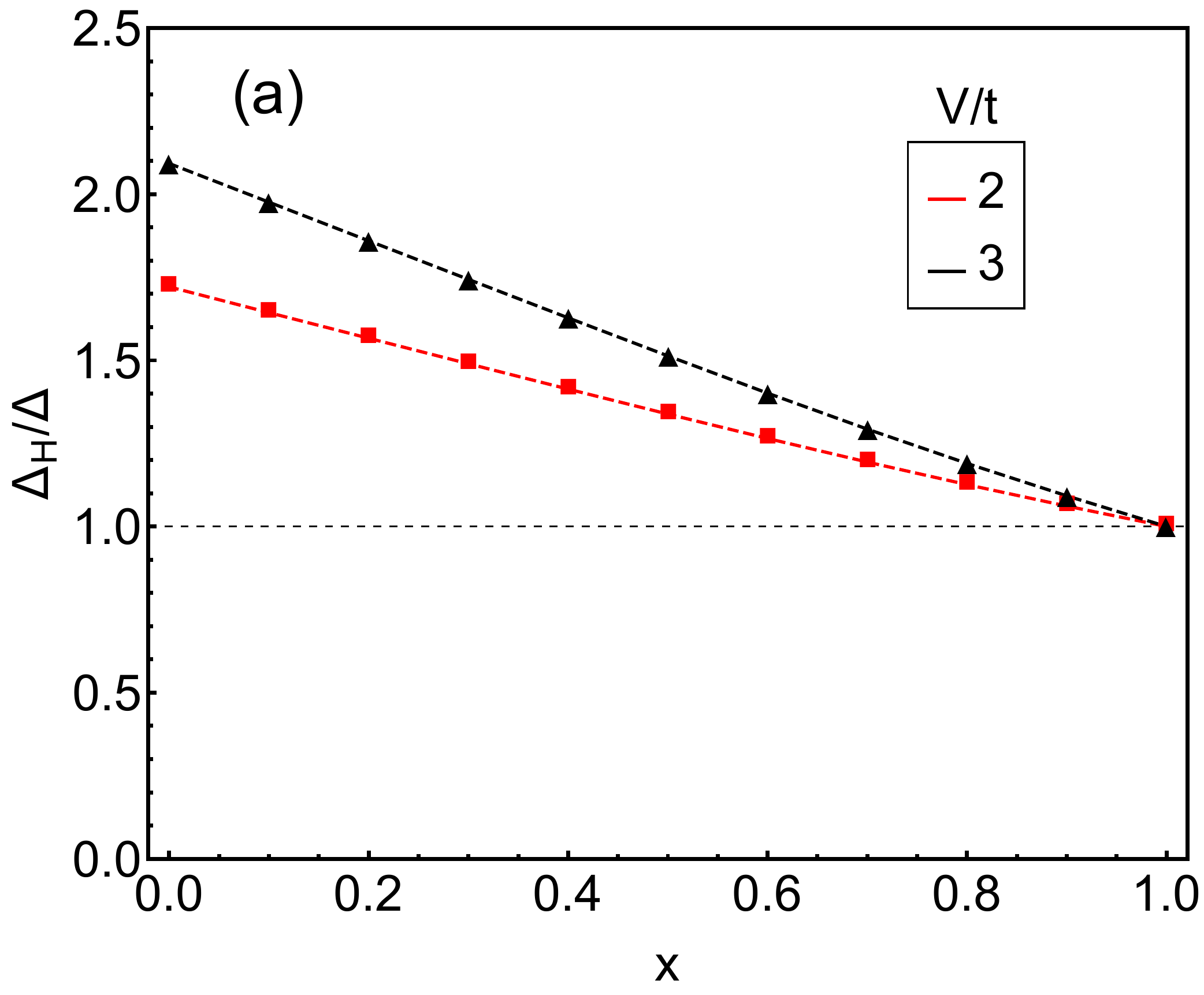} \hspace{40pt} \includegraphics[width=0.7\columnwidth]{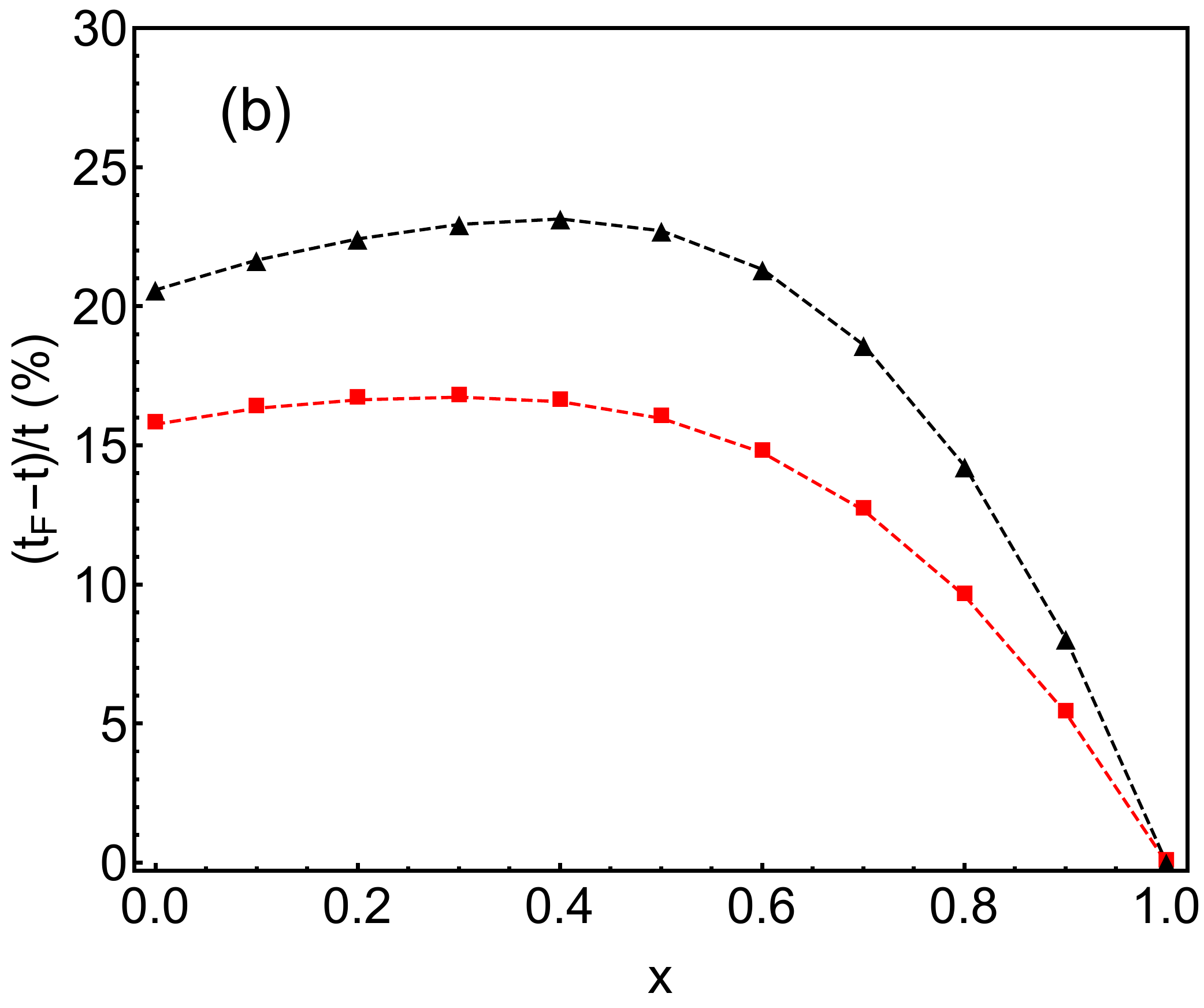}
\caption{a) $\Delta_{\rm H}/\Delta$ and b) the relative deviation $(t_{\rm F}-t)/t$ obtained for $\Delta=8t$ and $V/t=2,3$, as function of doping.}
\label{fig:Delta&t_HF}
\end{figure*}

Having described, within the RPA, the renormalization of the interaction kernel, we now turn to the renormalization of the single-particle Green function by interactions. These corrections come from the Hartree and Fock diagrams, and can be efficiently accounted for by a redefinition of the sublattice potential difference $\Delta \to \Delta_{\rm H}$ and of the NN tunneling amplitude $t \to t_{\rm F}$, with the value of $\Delta_{\rm H}$ and $t_{\rm F}$ determined self-consistently (see App.~\ref{app:HartreeFock}). In the large $N_f$ limit, the Hartree correction dominates and we find $\Delta_{\rm H} \simeq \Delta + 3V$ and $t_{\rm F} \simeq t$. Away from this analytical limit, we obtain $\Delta_{\rm H}$ and $t_{\rm F}$ numerically. As an illustration, Fig.~\ref{fig:Delta&t_HF} shows these two quantities obtained for $\Delta=8t$ and $V/t=2,3$ as function of doping. We observe that $\Delta_{\rm H}$ approaches linearly its bare value $\Delta$ as $x\to 1$, where the Hartree-Fock corrections vanish since both the $A$ and $B$ are completely filled. In addition, we find a non-zero doping-dependent correction to the hopping amplitude, $t_{\rm F}$, up to 20-25$\%$ of its bare value, $t$, for $V=3t$.

\subsection{Pairing in dilute limit}

The argument given in Sec.~\ref{sec:InteractionExpansion} can be repeated to find that the dilute effective model in the RPA regime also has $g_2 = 0$ and $g_1 = 2g_0$, such that it is described by Eq.~\ref{eq:ContinuumContinuumG0only}. Thus, the sign of $g_0$ determines the nature of interaction. This coefficients takes the form 
\begin{equation} \label{eq:valleysingletsRPA}
g_0 = \Gamma_{B,B}^{\rm RPA} (0) = \frac{9V^2}{N_f} \frac{\chi}{1+6V\chi} , 
\end{equation}
and indeed describe an attractive interaction since $\chi <0$. 

Our different analytical methods prove the presence of an effective attraction and the formation of bound electronic pairs in all the regimes highlighted in Fig.~\ref{fig:DomainValidity} for low enough densities. The reliability of our approach can be demonstrating by comparing the result of all three approaches in the joint region of applicability. It is clear that Eq.~\ref{eq:valleysingletsRPA} reduces to the results of the interaction expansion when $V\ll \Delta$, since they include the exact same diagrams. We now compare the RPA and HE results assuming both $N_f \gg 1$ and $t \ll\Delta$. In that case, the Hartree-Fock coefficient read $\Delta_{\rm H} \simeq \Delta + 3V$ and $t_{\rm F} \simeq t$ (App.~\ref{app:HartreeFock}) and the conduction band is almost flat $\varepsilon_{k,+} \simeq \Delta_{\rm H} / 2$. Finally, using $\sum_{{k}} |f({k})|^2 = 3N$, the RPA result can be written as  
\begin{equation}
g_0 \simeq - \frac{27 t^2 V^2}{N_f \Delta_{\rm H}^3} = - \frac{27 t^2 V^2}{N_f (\Delta+3V)^3} ,
\label{eq:g0}
\end{equation}
which perfectly agrees with the HE result of Eq.~\ref{eq:valleysingletenergyKE}, because $P \simeq (\Delta + 3V)^3$ when $N_f \gg 1$. This provide a stringent test of our methods, and rigorously demonstrate the presence of pairing between doped charge in our model.

\section{Numerical results} \label{sec:Superconductivity}

To go beyond infinitesimal doping and reach regimes where none of the perturbative parameters $t/\Delta$, $V/\Delta$ or $1/N_f$ are small, we now perform extensive numerical simulations of the original model Eq.~\ref{eq:OriginalSpinlessModel}. We include Hartree-Fock corrections and the repulsive interaction screened by particle-hole excitations described by $\Gamma^{\rm RPA}(q)$ for $N_f=1$.

The method then assumes the superconductivity to be driven by the charge fluctuations and is analogous to the one that two of us used recently for studying the origin of superconductivity in the twisted bi- and tri-layer graphene~\cite{Ceae2107874118,phong_cm21}, as well in the rhombohderal tri-layer graphene~\cite{cea_cm21}. Within this framework, the linearized gap equation is: 
\begin{align}\label{eq:linearized_gap}
\Delta_{n,k}&=-\frac{1}{N}\sum_{k',n'} V_{n',k'}^{n,k} \frac{\tanh\left(\beta\xi_{n',k'}/2\right)}{2\xi_{n',k'}} \Delta_{n',k'},  \\
V_{n',k'}^{n,k}&= \sum_{a,b=A,B} \Gamma^{\rm RPA}_{a,b}(k-k')  \Psi^{a,*}_{k,n} \Psi^{a}_{k',n'} \Psi^{b}_{k,n} \Psi^{b,*}_{k',n'} . \nonumber
\end{align}
For convenience, we rewrite the Eq.~\ref{eq:linearized_gap} in terms of an hermitian operator, as:
\begin{equation}\label{eq:linearized_gap_herm}
\tilde{\Delta}_{n,k}=-\sum_{k',n'}
\tilde{\Gamma}_{(n,k),(n',k')}
\tilde{\Delta}_{n',k'},
\end{equation}
where:
%\begin{widetext}
\begin{align} \label{eq:kernelnumerics}
\tilde{\Delta}_{n,k}&\equiv
\Delta_{n,k}\times
\sqrt{ \frac{\tanh\left(\beta\xi_{n,k}/2\right)}{2\xi_{n,k}}}
\\
\tilde{\Gamma}_{(n,k),(n',k')}&=
\frac{1}{N}
\sqrt{ \frac{\tanh\left(\beta\xi_{n,k}/2\right) \tanh\left(\beta\xi_{n',k'}/2\right)}{4 \xi_{n,k} \xi_{n',k'}}} V_{n',k'}^{n,k} . \nonumber % \times \sum_{a,b=A,B}\Gamma^{\rm RPA}_{a,b}(k-k') \Psi^{a,*}_{k,n}\Psi^{a}_{k',n'} \Psi^{b}_{k,n}\Psi^{b,*}_{k',n'}.
\end{align}
%\end{widetext}

The critical temperature, $T_c$, is defined as the largest value of $T$ such that the kernel $\tilde{\Gamma}$ has the eigenvalue 1, in order for Eq.~\ref{eq:linearized_gap_herm} to have solutions. The corresponding eigenvector provides the symmetry of the superconducting order parameter. In general, Eq.~\ref{eq:linearized_gap_herm} admits both symmetric ($A_1$ irrep of the lattice point group symmetry) and antisymmetric ($A_2$ irrep) solutions upon exchanging $k \rightarrow -k$. However, only the latter satisfies the Pauli's principle imposed by the statistics of spinless fermions. The $A_1$ solutions being unphysical, we discard them by projecting Eq.~\ref{eq:linearized_gap_herm} on the $A_2$ subspace:
\begin{equation}\label{eq:linearized_gap_herm_A2}
\tilde{\Delta}_{n,k}=-\sum_{k',n'}
\tilde{\Gamma}^{A_2}_{(n,k),(n',k')}
\tilde{\Delta}_{n',k'},
\end{equation}
with:
\begin{equation}
\tilde{\Gamma}^{A_2}_{(n,k),(n',k')}\equiv\frac{
\tilde{\Gamma}_{(n,k),(n',k')}-
\tilde{\Gamma}_{(n,k),(n',-k')}}{2}.
\end{equation}

\begin{figure}
\centering
\includegraphics[width=0.9\columnwidth]{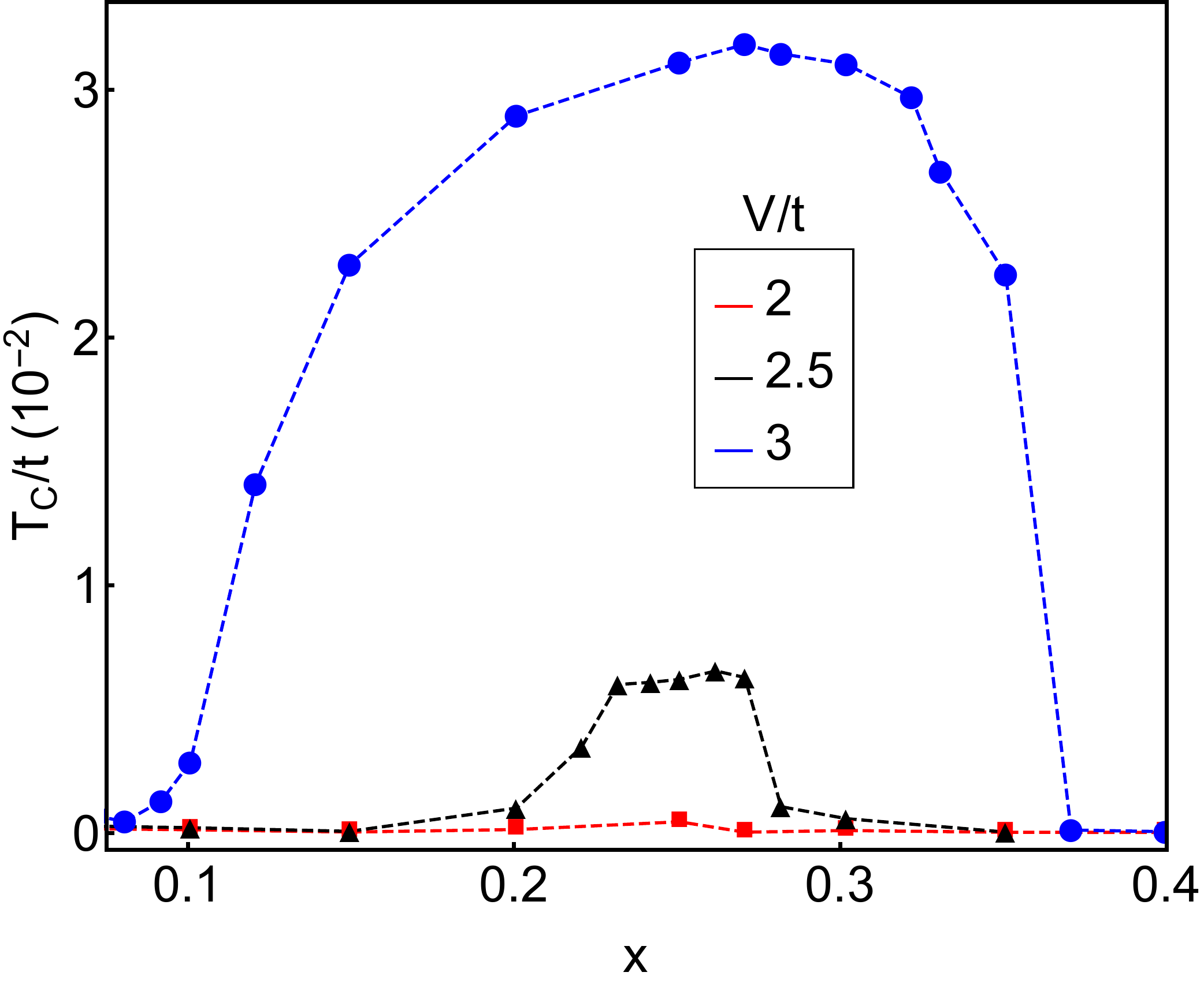}
\caption{
Values of the critical temperature as function of the doping, obtained by solving numerically the Eq. \eqref{eq:linearized_gap_herm_A2}, for $\Delta=8t$ and $V/t=2,2.5,3$. We use $N=2700$ unit cells and include the HF corrections to the band structure and the eigenfunctions. }
\label{fig:RPA_KL_Tc}
\end{figure}

We now present our numerical solutions of Eq.~\ref{eq:linearized_gap_herm_A2}, with the band structure and the eigenfunctions accounting for the HF corrections (see Fig.~\ref{fig:Delta&t_HF}). We sample the BZ with $N=2700$ points, set $\Delta=8t$ and consider different values of $V$ upon varying the doping $x$. Fig.~\ref{fig:RPA_KL_Tc} shows $T_c$ as function of $x$ for $V/t=2,2.5,3$. The critical temperature displays a non-monotonic behavior, and reaches a maximum at an optimal doping $x\simeq 0.25$. As expected when the superconductivity is driven by electronic interactions, $T_c$ increases with the interaction's strength $V$, reaching values up to $\sim 3-3.5\times10^{-2}t$ for $V/t=3$.

\begin{figure}
\centering
\includegraphics[width=1.1\columnwidth]{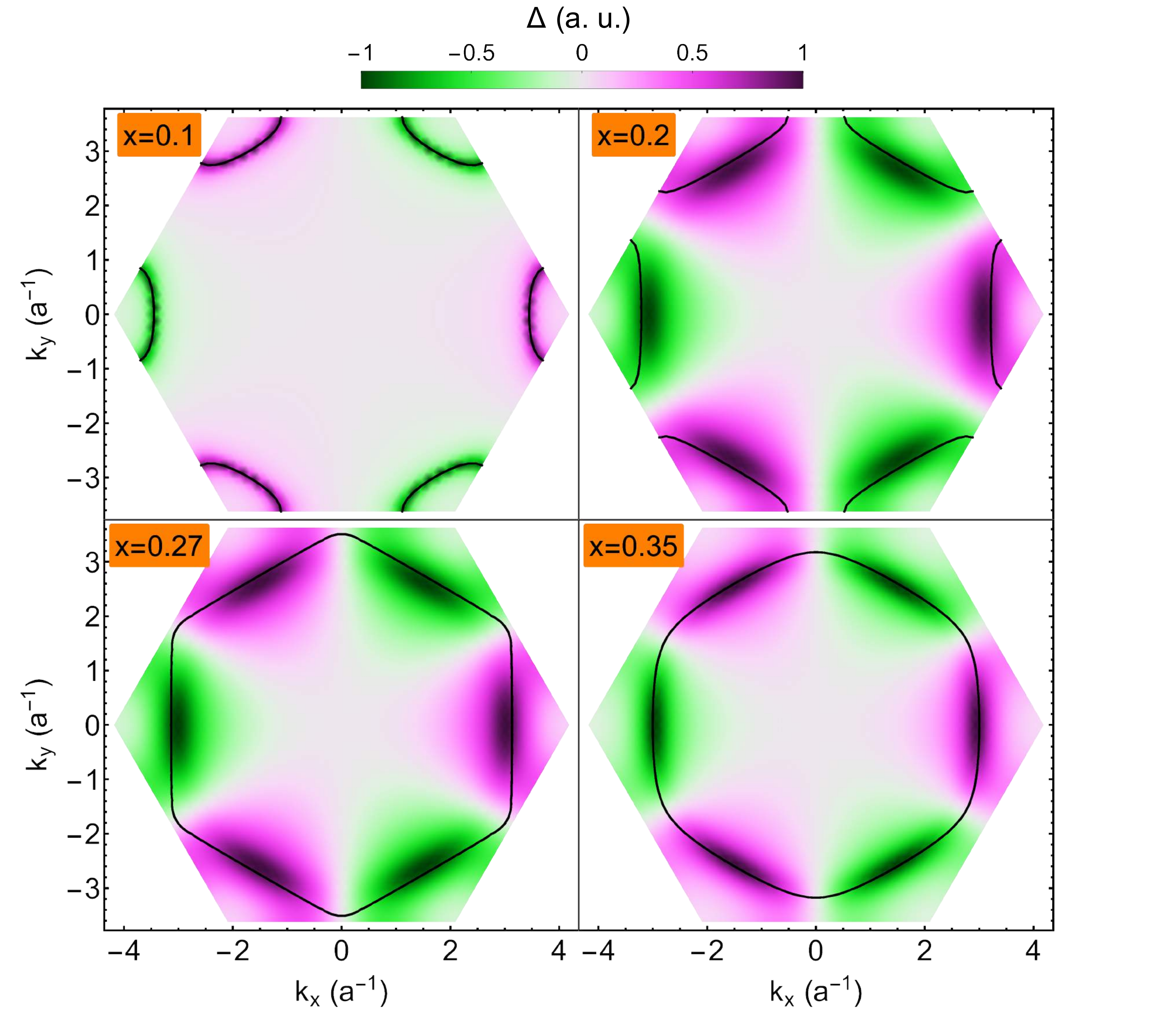}
\caption{
Symmetry of the superconducting order parameter in the BZ, obtained for $\Delta=8t$, $V=3t$ and four different values of doping. The continuum black lines represent the Fermi surface.
}
\label{fig:RPA_KL_OP}
\end{figure}

Fig.~\ref{fig:RPA_KL_OP} shows the symmetry of the superconducting order parameter in the BZ, obtained for $\Delta=8t$, $V=3t$ and four different values of the doping. The continuum black lines represent the Fermi surface. Comparing the Figs.~\ref{fig:RPA_KL_OP} and~\ref{fig:RPA_KL_Tc}, we note that $T_c$ increases with the size of the Fermi surface. The latter is a non-monotonic function of the doping, increasing with $x$ up to $x\sim 0.26$ and then shrinking around the centre of the BZ. We also remark that $T_c$ falls down abruptly when $x \gtrsim 0.26$, which can be understood from the emergence of nodes in the superconducting order parameter, as can be seen in Fig.~\ref{fig:RPA_KL_OP}. 

\section{Discussion} \label{sec:Discussionconsequences}

\subsection{Interband polarization}

Our analytic arguments establishing the presence of attraction and the formation of bound pairs in the low carrier density limit (see Sec.~\ref{sec:HybridizationExpansion}) offer a new perspective from many-body perturbation theory on the fully electronic mechanism for superconductivity introduced in Ref.~\cite{CrepelFu_SCfromRepulsion}. More precisely, the pairing interaction emerges from the polarization bubble Eq.~\ref{eq:bubbleSpecificpairing}, which is dominated by interband contributions (see Sec.~\ref{sec:InteractionExpansion}). Therefore the emergence of superconductivity in our model is due to interband polarization. We further substantiate this statement in App.~\ref{app:diagrams}, where we present analyse the diagrams leading to attractive interaction at small momenta in more details.

The interband process leading to pairing interaction corresponds to a virtual state having one hole in the valence band and one additional conduction electron. This perfectly connects to the ``three-particle mechanism'' introduced in Refs.~\cite{CrepelFu_SCfromRepulsion,CrepelFu_TripletSC}, where the pairing between two conduction electrons is mediated by a third electron in the valence band undergoing a virtual interband transition. Importantly, this mechanism differs from other interband mechanism of superconductivity relying on virtual {\it pair} scattering of {\it two} electrons from the Fermi surface to distant or incipient bands~\cite{chen2015electron,dong2021activating}, as described by the ladder diagram of Eq.~\ref{eq:FiveKLDiagrams} rather than the bubble diagram.

\begin{figure}
\centering
\includegraphics[width=\columnwidth]{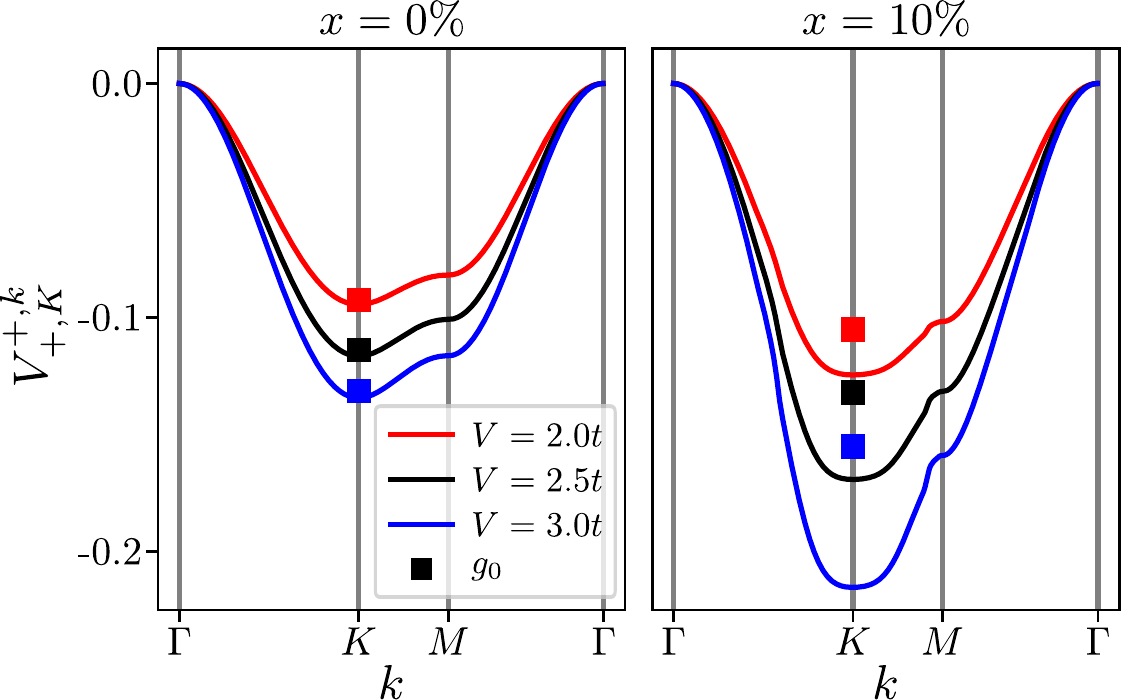}
\caption{
The effective Cooper interaction $V_{n',k'}^{n,k}$ (see Eq.~\ref{eq:linearized_gap}) in the upper band $n=n'=+$ is attractive, as shown here using the incoming momentum $k'=K$ as reference, with $\Delta=8t$ and two different doping concentrations. Squares show the analytical prediction $g_0$ (Eq.~\ref{eq:g0}) obtained by neglecting intraband contributions.
}
\label{fig:IntraInterBand}
\end{figure}

To get more insights into the effective attraction between electrons leading to the formation of the Cooper pairs, we study in more the details the effective Cooper interaction $V_{n',k'}^{n,k}$ in the upper band $n=n'=+$ obtained within RPA, see Eq.~\ref{eq:linearized_gap}. For illustration purposes, we fix the incoming vector $k'=K$ at the center of the Fermi sea, and plot the interaction strength as a function of $k$ in Fig.~\ref{fig:IntraInterBand} in the dilute limit $x\simeq 0\%$, obtained by fixing $\mu$ at the valence band bottom, and for a substantial doping concentration $x=10\%$. First and foremost, we observe that the effective interactions in the upper band are attractive, as expected from our analytical and numerical results of Sec.~\ref{sec:HybridizationExpansion}~-~\ref{sec:Superconductivity}. At infinitesimal doping $x \simeq 0\%$, obtained by fixing $\mu$ at the valence band bottom, we observe that this attraction is entirely due to interband polarization effects. Indeed, it almost perfectly agrees with our analytical prediction for $g_0$ in Eq.~\ref{eq:g0} when the momentum transfer $|k'-k|\to 0$ (squares in Fig.~\ref{fig:IntraInterBand}). 
On the other hand, for larger doping concentrations, exemplified here with $x=10\%$, the intraband susceptibilities become non-negligible and start to strongly renormalize the effective RPA interactions. These intraband contributions are similar to those appearing in the standard Kohn-Luttinger mechanism. For the values chosen here, they increase the strength of the attractive potential by up to $30 \%$ compared to the pure interband polarization effects. The increase of $T_c$ for $x = 0.1 - 0.25$ observed in Fig.~\ref{fig:RPA_KL_Tc} is thus due to a combined effects of the interband and intraband polarization.

\subsection{Contrast with Kohn-Luttinger mechanism}

It is worth to compare the mechanism for superconductivity studied here with the traditional Kohn-Luttinger mechanism. From a technical viewpoint, we highlighted in Sec.~\ref{sec:InteractionExpansion} that the susceptibilities consist of two different contributions: intraband, englobed by the standard Kohn-Luttinger theory, and interband that greatly dominate at low doping (see Fig.~\ref{fig:IntraInterBand}). The non-zero interband polarization has important physical consequences.

First, our screened interaction, once projected near the Fermi energy, is attractive for small momentum transfers $q \ll k_F$. This is not the case in the KL mechanism, which relies on screening processes near $q=2k_F$. These high momentum oscillations of the screened Coulomb potential can, at sufficiently large distances, reach negative values. This produces an effective attraction at momentum transfer close to $2k_F$ and pairing in high orbital angular channel, which often leads to nodal superconducting order parameter. Here in contrast, the interband polarization bubble is already attractive at small $q$ (see also App.~\ref{app:diagrams}), which enables a nodeless order parameter at small densities. 

Second, the KL mechanism  requires a finite carrier density leading to a Fermi surface~\cite{maiti2013superconductivity,kagan2014kohn, nandkishore2014superconductivity,lin2018kohn, NLC12,GS19,YIF19} to produce screened interaction. To put it another way, Kohn-Luttinger mechanism does not produce any bound state when only two electrons are present and subject to mutual repulsion. In contrast, our system consists of a completely filled band of ``core'' electrons, which induces the pairing interaction between doped electrons in the conduction band. Remarkably, we find that two doped electrons already attract and form a bound state, and this attractive interaction is the result of interband electronic effect in a band insulator with repulsive bare interaction. This novel interband mechanism distinguishes our work from Kohn-Luttinger mechanism, and produces robust pairing at {\it infinitesimal} doping.

\subsection{Role of correlated hopping} \label{ssec:Correlatedtunneling}

In the real-space picture offered by the hybridization expansion (Sec.~\ref{sec:HybridizationExpansion}), the emergence of superconductivity crucially relies on correlated hopping processes of doped electrons -- the $\lambda$ term Eq.~\ref{eq:StrongCouplingHamilto}. Historically, correlated hopping was discussed in connection with superconductivity in the cuprates~\cite{HM90,MH90} (see also~\cite{SV34,VK79}). However, the correlated hopping term discussed for cuprates differs from the $\lambda$ term that emerges in the low-energy limit of our model. The former describes electrons hopping between two sites, one of which being already occupied by another electron of opposite spin. This process necessarily involves double occupation and therefore is strongly suppressed by large on-site repulsion $U$. In contrast, our $\lambda$ term describes electron hopping between two sites in the presence of another electron occupying a third site nearby, a process that remains active and leads to attractive interactions even at large $U$~\cite{CrepelFu_TripletSC}.

These density-dependent hopping terms manifest themselves as an increase of the effective tunneling strength for doped charges as a function of their density. This behavior is also captured by the sublattice potential in the Hartree approximation $\Delta_{\rm H} \approx \Delta + 3V(1-x)$ (see Fig.~\ref{fig:Delta&t_HF}). Notably, the interaction-induced renormalization of $\Delta_{\rm H}$ makes the effective hopping amplitude doping-dependent: 
\begin{equation}
t_{\rm eff} = \frac{t^2}{\Delta_{\rm H}} \approx \frac{t^2}{ \Delta + 3 V} + \frac{ 3 t^2 V x }{ (\Delta + 3 V )^2}  .
\label{hoppinghartree}
\end{equation}
%In Sec.~\ref{subsec:tbg}, we shall show that 
This effect is captured by the correlated hopping term in the effective Hamiltonian after downfolding, and it is essential for the emergence of superconductivity in our model.

\section{Relevance for twisted bilayer graphene}
\label{subsec:tbg}

%\subsection{Analogy}

Throughout this paper, we have highlighted that superconductivity at low doping above integer filling can arise from repulsive interactions in multiband systems due to a nonzero virtual interband polarization. The cause of the attractive interactions is the presence of correlated hopping terms for doped electrons, which can be identified by a filling-dependent Hartree potential $\Delta_{\rm H}$.

Similar correlated hopping terms have also been identified observed twisted bilayer graphene (TBG), where the Hartree correction experiences significant changes as a function of band filling~\cite{Guinea_pnas18}. Drawing an analogy between the two systems, we expect the correlated hopping terms to emerge in the low-energy theory of twisted bilayer graphene and to be responsible for superconductivity.

To place this analogy on firmer grounds, we now examine and compare effective interactions in TBG and in our model using the RPA of Sec.~\ref{sec:RandomPhaseApproximation}, and show the evident similarity between the two.

%\subsection{Emergence of attractive interactions}

Let us first briefly summarize our results for the spinless model Eq.~\ref{eq:OriginalSpinlessModel}. The renormalization of the sublattice potential difference is the only inhomogeneous on-site potential that respect the the translational symmetry of the system. It takes the form $V_{\rm H}(r) = \Delta_{\rm H} [\rho_A (r) - \rho_B(r)]/2$, with $\rho_{A/B} (r)$ the density on each orbitals in the unit cell located at $r$. Because of this orbital structures, all quantities involved in the renormalization of the scattering vertex, such as the polarizability $\chi_{a,b}$, are promoted to matrix form with orbital indices (see Eq.~\ref{eq:RPAsuscdef}).

In TBG, the Hartree term contains all possible inhomogeneous on-site potentials with the periodicity of the moir\'e unit cell.  Its Fourier components are
\begin{equation}
V_{\rm H} ( r ) = \sum_{G} \frac{2 \pi e^2}{| G |} \rho_{G} (r) ,
\end{equation}
where $G \neq 0$ is a moiré reciprocal lattice vector, and $\rho_{G}$ is a Fourier component of the total charge density. The latter can be written as
\begin{equation}
\rho_{G} (r) = \sum_{k , n \in occ.} \braOket{ k , n}{ e^{i G r} }{ k , n } , 
\end{equation}
where contraction over the layer and sublattice indices is implied in the expectation value of the single-particle wavefunctions $\ket{k,n}$, which are now four component spinors.

Hence, the Hartree potential can be written in terms of the $\rho_G$, in the same manner that the Hartree potential of the model discussed in Sec.~\ref{sec:Modelandmethods} can be writen in terms of $\rho_{A/B}$, the role of the sublattice index being replaced by the reciprocal lattice vectors. The analog goes deeper. In the honeycomb model the two sublattices are connected by electron tunneling $t$, and similarly,  electron states of graphene at wavevectors $k$ and $k+G$ are connected by the interlayer tunneling. In both cases, single-particle bands are formed by the hybridization between multiple components within a unit cell.

Furthermore, as the number of flavors is larger in TBG (spin, layer, sublattice), the RPA treatment becomes more and more justified. Following the reasoning of Sec.~\ref{sec:RandomPhaseApproximation}, we find that the susceptibility and RPA vertex are now promoted to matrices, $\chi_{G , G'} (q)$ and $\Gamma_{G , G'}^{\rm RPA} (q)$, indexed by the moiré reciprocal lattice vectors, in the same way that the repulsive potential and the susceptibility of the model in Sec.~\ref{sec:Modelandmethods} can be written as a matrix using sublattice indices. 

\begin{figure}
\centering
\includegraphics[width=\columnwidth]{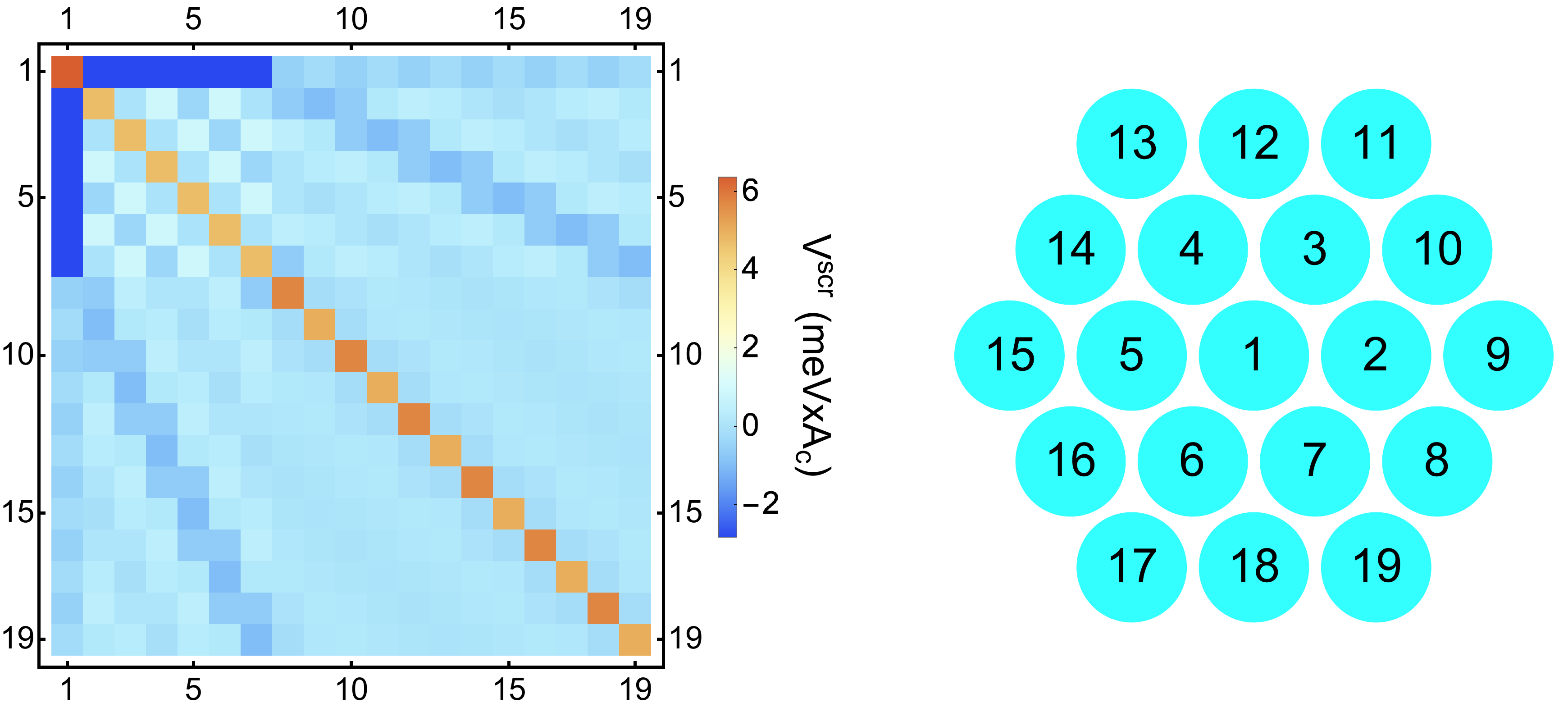}
\caption{
Some matrix elements $\Gamma_{G , G'}^{\rm RPA} ( q \rightarrow 0 )$ of the screened potential for twisted bilayer graphene, obtained for the twist angle $\theta=1.05^\circ$ and for $x=-1$ electrons per moir\'e unit cell. The index $G$ denotes moiré reciprocal lattice vectors, as defined on the right.}
\label{fig:veff}
\end{figure}

To compare the physics of the two models, we plot in Fig.~\ref{fig:veff} the RPA vertex $\Gamma_{G , G'}^{\rm RPA} ( q \rightarrow 0 )$ obtained for TBG near the magic angle, at filling $x=-1$ (measured with respect to charge neutrality). % where interband effects dominate (see Ref.~\cite{Ceae2107874118} for methods). 
We observe that the {\it off-diagonal} elements of this kernel, not initially present in the interaction, are now negative, signaling attractive interactions. 

The analogy between the spinless model Eq.~\ref{eq:OriginalSpinlessModel} and the interacting model of TBG consolidates. In both cases, the matrix which defines the screened RPA vertex acquires new elements, which are not present in the bare interaction, but emerge from {\it interband} polarization. 
%The effect of correlated hopping is to reverse the sign of these new coefficient, in spite of the bare original repulsive interaction. 
Furthermore, the order parameter for TBG induced by these new attractive terms shows few or no sign changes throughout the mini Brillouin Zone of the moir\'e structure~\cite{Ceae2107874118}, as in the spinless model described here~\cite{CrepelFu_SCfromRepulsion,CrepelFu_TripletSC}.

We emphasize that attractive part of the interaction in TBG, shown in Fig.~\ref{fig:veff}, is largely due to interband processes. The essential role of interband contribution to superconductivity is further testified by considering an approximation method that discards it. This approximation, known as the "flat-metric condition", is convenient for the analytical study of the effect of interactions in TBG, and has been extensively applied in Refs.~\cite{TBG_I,TBG_II,TBG_III,TBG_IV,TBG_V,TBG_VI}. As described in App.~\ref{app:FlatMetric}, we show that this approximation however leads to a purely repulsive interaction $V_{c,k'}^{c,k}$ in both TBG and our honeycomb model, thus failing to capture the emergence of superconductivity. %Hence, washing out the interband contributions and the correlated hopping terms is too restrictive to explain the results obtained throughout this paper, and in TBG. 

\section{Conclusion} \label{sec:conclusion}

In this article, we comprehensively studied one of the simplest electronic mechanism for superconductivity in doped band insulators. Using a minimal two-band model for illustration, we firmly established the emergence of attractive interactions between doped charges with the help of three analytically controlled methods, whose results perfectly agree in their overlapping domains of validity. 
All three methods convey a common understanding for the origin of the effective attraction: it arises from the coupling of conduction electrons pairs to virtual excitons. %Only our description of this coupling differs between the various parameter regimes. 
In  hybridization expansion in the real-space picture, this effective attraction is described by a correlated hopping term in the effective Hamiltonian for doped electrons, while in the RPA framework based on the $k$-space band picture, it arises from interband contribution to the particle-hole polarization bubble. 

The pairing interaction described here shows a feature not commonly found in models of superconductivity from bare repulsive interactions: the screened interaction, projected onto the states near the Fermi energy, is attractive for small momentum tranfer~\cite{note}, $| q | \ll k_F$.

We have characterized the superconducting order parameter using full fledged numerical calculations. The small momentum attraction mentioned above translates into a nodeless order parameter at the Fermi surface pockets at the corners of the Brillouin Zone, the points $K$ and $K'$. We found that the critical temperature $T_c$ exhibits a dome as a function of doping, whose maximum reaches up to $3\%$ of the nearest neighbor hopping amplitude. For a realistic value $t=0.1$eV, this already leads to a substantial $T_c\simeq 35$K. This maximum appears near $x = 1/4$, above which the superconducting gap acquires nodes, leading to a reduction of $T_c$.

Finally, we highlighted some consequences of our work regarding the theoretical modeling of interactions in twisted bilayer graphene. In particular, we showed the crucial role of correlated hopping terms arising from interband polarization in the understanding of superconductivity. This hints that approximation schemes discarding their effects will probably also underestimate the strength of attractive interaction -- if they find any. Remarkably, the structure of the interaction matrix in our simple model and in twisted bilayer graphene are analogous, with attractive off-diagonal driving the transition to a superconducting state. This shows the importance of the electronic mechanism of superconductivity studied in this article for the understanding of more complex systems.

\section{Acknowledgements} 

The work of FG and TC was supported by funding from the European Commission, under the Graphene Flagship, Core 3, grant number 881603, and by the grants NMAT2D (Comunidad de Madrid, Spain),  SprQuMat and SEV-2016-0686, (Ministerio de Ciencia e Innovación, Spain).
The work at Massachusetts Institute of Technology is funded by the Simons Foundation through a Simons Investigator Award. VC gratefully acknowledges support from the MathWorks fellowship. LF is partly supported by the David and Lucile Packard Foundation.

\bibliography{BiblioRPASC}

\begin{thebibliography}{62}
\bibitem [{not()}]{note}%
  \BibitemOpen
  \href@noop {} {}\bibinfo {note} {Ref.\cite{HWQM21} also obtains, for certain
  parameter values, an attractive interaction for $\vec{q} \rightarrow 0$. This
  result is achieved, as in our work, because of a large contribution from the
  sublattice polarizability in a model of spinless electrons. In the model
  studied here the sublattice degree of freedom is the only internal index
  available. The extension of the results to twisted bilayer graphene is based
  on a susceptibility where charge fluctuations within the unit cell, labelled
  by reciprocal lattice vectors, are crucial, see
  Section\ref{subsec:tbg}.}\BibitemShut {Stop}%
\end{thebibliography}

\appendix

\newpage

\section{Conventions for the lattice} \label{app:HoneycombLattice}

For completeness, we provide in this appendix one possible parametrization of the honeycomb lattice and of its first Brillouin zone. 
The explicit expressions given below ease the analytical evaluation of the $g$-coefficients appearing in the continuum model of Sec.~\ref{sec:HybridizationExpansion}, for instance. 

\begin{figure}
\centering
\includegraphics[width=0.9\columnwidth]{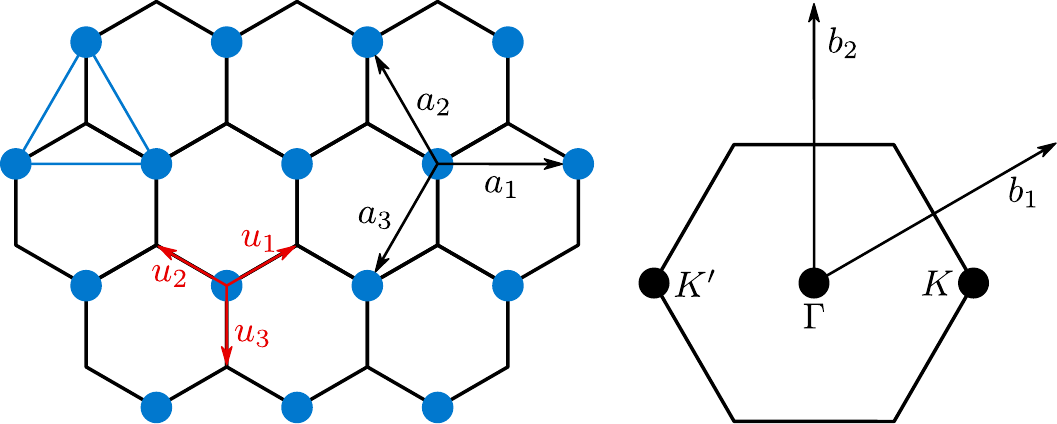}
\caption{On the left, primitive lattice vectors $a_{j=1,2,3}$ and nearest-neighbor vectors $u_{j=1,2,3}$ on the honeycomb lattice. 
Highlighted in blue is the triangular $B$-lattice and the upper triangle centered on a $A$ site summed over in the effective KE model (see Eq.~\ref{eq:StrongCouplingHamilto}). 
On the right, first Brillouin zone and high smmetry points $\Gamma$, $K$ and $K'$. 
}
\label{fig:HoneycombLattice}
\end{figure}

We define, as shown in Fig.~\ref{fig:HoneycombLattice}, $(a_1, a_2)$ the two primary lattice basis vectors; $a_3 = -a_2-a_1$ the third primitive lattice vector; $a = |a_1| = |a_2| = |a_3|$ the lattice constant; $(u_1, u_2, u_3)$ the three nearest-neighbor vectors pointing toward $A$ sites; $(b_1, b_2)$ the reciprocal lattice basis vectors; $\Gamma$ the center of the Brillouin zone; and $(K, K')$ its two nonequivalent corners. 
Their expression and properties can be straightforwardly obtained from Fig.~\ref{fig:HoneycombLattice}. 
We also gather for convenience some useful relations 
\begin{equation}
\begin{array}{c||c|c|c|c|c|c} 
v & a_1 &  a_2 &  a_2 &  u_1 &  u_2 &  u_3
\\\hline\hline
K \cdot v & - 2 \pi/3 & 4\pi/3 & -2\pi/3 & 0 & 2\pi/3 & - 2 \pi/3 \\\hline
K' \cdot v & 2 \pi/3 & -4\pi/3 & 2\pi/3 & 0 & -2\pi/3 &  2 \pi/3 
\end{array} \, \, ,
\end{equation}
which show -- for instance -- that $f(K)=f(K')=0$, or $\Lambda_{K,K}^{\rm KE}(0) = \Lambda_{K,K'}^{\rm KE}(0) = 7V_f/2-9\lambda$ and $\Lambda_{K,K}^{\rm KE}(K) = -V_f$ using Eq.~\ref{eq:StrongCouplingHamilto}.

\section{Projection of interaction} \label{app:EffectiveContinuum}

We now detail the projection of the interaction term of HE effective model near the band bottom, used to derive the continuum model Eq.~\ref{eq:ContinuumInteractionTerm} of the main text. We express the interacting part of our effective Hamiltonian as
\begin{equation}
H_{\rm int} = \frac{1}{N} \sum_{\substack{k,k',q\\\sigma,\sigma'}} \Lambda_{k,k'}^{\rm HE}(q) c_{k+q,B,\sigma}^\dagger c_{k',B,\sigma'}^\dagger c_{k'+q,B,\sigma'} c_{k,B,\sigma} ,
\end{equation}
and Fourier transform Eq.~\ref{eq:StrongCouplingHamilto} to get the interaction coefficients in the HE regime ($t\ll\Delta$)
\begin{widetext} \begin{equation} \label{appeq:ScattVertexKE}
\Lambda_{{k},{k'}}^{\rm HE} ({q}) = V_f \left[ \frac{3}{2} + \sum_{j=1}^3 \cos(q\cdot a_j) \right]   + \lambda \sum_{j=1}^3 2\cos (k \cdot a_j) + 2\cos [(q+k) \cdot a_j] + [ e^{i({q}\cdot{a}_j-{k}\cdot{a}_{j+1})} + e^{-i({q}\cdot{a}_j-{k}\cdot{a}_{j-1})} ]  . 
\end{equation} \end{widetext}
The projection restrict the sum to contributions where the four fermionic operators lie close to the $K$ of $K'$ valleys. Amongst these contributions, we can also discard the ones that cannot satisfy momentum conservation, which either involves three electrons in the same valley (\textit{e.g.} $\psi_{K}^\dagger \psi_{K}^\dagger \psi_{K} \psi_{K'}$) or scatter pair between the two valleys (\textit{e.g.} $\psi_{K}^\dagger \psi_{K}^\dagger \psi_{K'} \psi_{K'}$). 
We get 
\begin{equation} \begin{split}
H_{\rm int} \simeq & \frac{1}{N} \! \sum_{\substack{k,k',q\\\sigma,\sigma'}} \! [ \Lambda_{K,K}^{\rm HE}(0) \psi_{k+q,\sigma,K}^\dagger \psi_{k',\sigma',K}^\dagger \psi_{k'+q,\sigma',K} \psi_{k,\sigma,K} \\
& + \Lambda_{K',K'}^{\rm HE}(0) \psi_{k+q,\sigma,K'}^\dagger \psi_{k',\sigma',K'}^\dagger \psi_{k'+q,\sigma',K'} \psi_{k,\sigma,K'} \\
& + \Lambda_{K,K'}^{\rm HE}(0) \psi_{k+q,\sigma,K}^\dagger \psi_{k',\sigma',K'}^\dagger \psi_{k'+q,\sigma',K'} \psi_{k,\sigma,K}  \\
& + \Lambda_{K',K}^{\rm HE}(0) \psi_{k+q,\sigma,K'}^\dagger \psi_{k',\sigma',K}^\dagger \psi_{k'+q,\sigma',K} \psi_{k,\sigma,K'}  \\
& + \Lambda_{K,K}^{\rm HE}(K) \psi_{k+q,\sigma,K'}^\dagger \psi_{k',\sigma',K}^\dagger \psi_{k'+q,\sigma',K'} \psi_{k,\sigma,K}  \\
& + \Lambda_{K',K'}^{\rm HE}(K') \psi_{k+q,\sigma,K}^\dagger \psi_{k',\sigma',K'}^\dagger \psi_{k'+q,\sigma',K} \psi_{k,\sigma,K'} ] .
\end{split} \end{equation}
Using Hamiltonian's hermiticity and time-reversal invariance (or the explicit expression above), we get the relations $[\Lambda_{k,k'}(q)]^* = \Lambda_{k+q,k'+q}(-q) = \Lambda_{-k,-k'}(-q)$ allowing to equate some of the coefficients above, and to find
\begin{align} 
H_{\rm int} & \simeq \frac{1}{N} \sum_{\substack{q,\tau \\ \sigma,\sigma'}} \Lambda_{K,K}^{\rm HE}(0) \rho_{q,\sigma,\tau} \rho_{-q,\sigma',\tau} \\
& +  \frac{1}{N} \sum_{q,\sigma,\sigma'} 2 \Lambda_{K,K'}^{\rm HE}(0) \rho_{q,\sigma,K} \rho_{-q,\sigma',K'} \notag \\
& +  \frac{1}{N} \sum_{q,\sigma,\sigma'} \Lambda_{K,K}^{\rm HE}(K) [ \tau_{q,\sigma}^+  \tau_{-q,\sigma'}^- + \tau_{q,\sigma}^- \tau_{-q,\sigma}^- ] . \notag 
\end{align}
Using $\tau_\sigma^+ \tau_{\sigma'}^- + \tau_\sigma^- \tau_{\sigma'}^+ = [2\vec{\tau}_\sigma\cdot\vec{\tau}_{\sigma'} - (\rho_{K,\sigma} - \rho_{K',\sigma}) (\rho_{K,\sigma'} - \rho_{K',\sigma'})/2]$, we end up with the continuum model of Eq.~\ref{eq:ContinuumInteractionTerm} where the $g$-coefficient take the explicit form 
\begin{subequations} \begin{eqnarray}
g_0 &=& \Lambda_{K,K}^{\rm HE}(0) - \Lambda_{K,K}^{\rm HE}(K)/2 , \\ 
g_1 &=& 2\Lambda_{K,K'}^{\rm HE}(0) + \Lambda_{K,K}^{\rm HE}(K) , \\ 
g_2 &=& 2 \Lambda_{K,K}^{\rm HE}(K) ,
\end{eqnarray} \end{subequations}
which has been reproduced in Eq.~\ref{eq:InteractionExpansionGCoeff} concerning the interaction expansion -- the derivation being identical in that case. Using the explicit form of $\Lambda^{\rm HE}$, we find
\begin{equation}
g_2 = 0 , \quad g_1 = 2g_0 = 9(V_f - 2\lambda)  ,
\end{equation}
as announced in the main text.

\section{Formula for $T_c$} \label{app:FormulaTc}

We briefly reproduce the calculation of Refs.~\cite{miyake1983fermi,randeria1990superconductivity} leading to the explicit expression of $T_c$ in Eq.~\ref{eq:TcHybridizationExpansion}. When $N_f=1$, the continuum model of Eq.~\ref{eq:ContinuumContinuumG0only} is a two-component Fermi gas in two dimension with attractive interactions $g_0<0$, and a quadratic band dispersion with effective mass $m$. This theory hosts a bound state of energy $-\varepsilon_b$ determined as a pole of the two-body scattering $T$-matrix
\begin{equation}
\frac{1}{|g_0|} = \mathcal{D}_0 \int {\rm d} \varepsilon \; \frac{1}{\varepsilon_b + 2 \varepsilon} , 
\end{equation}
with $\mathcal{D}_0 = Am/(\pi\hbar^2)$ the constant density of state in two dimensions and $A$ the area of the unit cell in real space. Its solution gives $\varepsilon_b \propto W e^{-1/g}$ when $g$, given below Eq.~\ref{eq:TcHybridizationExpansion}, is large~\cite{CrepelFu_SCfromRepulsion}. 

We then solve the gap and number equations
\begin{subequations} \label{eq:AppGapNumberEquation}
\begin{eqnarray}
\frac{1}{|g_0|} &=& \mathcal{D}_0 \int_0^\infty {\rm d} \varepsilon \; \frac{\tanh(\beta E_k /2 )}{2E_k} , \\
2E_F &=& \int_0^\infty {\rm d} \varepsilon \;  \left[ 1 - \frac{\varepsilon - \mu}{E_k} \tanh \left( \frac{\beta E_k}{2} \right) \right] ,
\end{eqnarray}
\end{subequations}
where $E_k =  \sqrt{(\varepsilon_k-\mu)^2 + \Delta^2}$
for $T=0$ and $T = T_c$ by direct evaluation of the integrals above. At $T=0$, the number equation gives
\begin{equation}
\sqrt{\mu^2 + \Delta^2} + \mu = 2 E_F , 
\end{equation}
where $\mu$ is the chemical potential, while the gap equation can be recast as 
\begin{equation}
\int_0^\infty {\rm d} \varepsilon \; \left[ \frac{1}{2E_k} - \frac{1}{2\varepsilon + \varepsilon_b} \right] = 0 , 
\end{equation}
which is convergent and yields a second relation between $\mu$ and $\Delta$:
\begin{equation}
\sqrt{\mu^2 + \Delta^2} - \mu = \varepsilon_b .  
\end{equation}
Together, they give 
\begin{equation}
\mu = E_F - \varepsilon/2 , \quad \Delta = \sqrt{2 E_F \varepsilon_b} . 
\end{equation}

Because the two-dimensional density of state is constant, we can reproduce the standard BCS integrals without assuming any separation of scale between the UV cutoff $\varepsilon_{\Lambda}$ and the Fermi energy. This yields
\begin{equation}
k_B T_c = K \Delta \propto \sqrt{E_F \varepsilon_b} \propto \sqrt{E_F W} e^{-1/(2g)} ,
\end{equation}
as announced in the main text. Note that this equation applies to the dilute limit of all regime considered (KE, KL and RPA), with the $g_0$ respectively given by Eqs.~\ref{eq:valleysingletenergyKE},~\ref{eq:valleysingletenergyKL} and~\ref{eq:valleysingletsRPA}. 

For instance, let us highlight how the RPA linearized gap equation  Eq.~\ref{eq:linearized_gap} gets mapped onto Eq.~\ref{eq:AppGapNumberEquation} in the dilute limit. In this regime electrons mostly live near $\pm K$, and interact through small momentum transfers $q \simeq 0$ since $\Gamma^{\rm RPA}(\pm \vec{K}) = 0$. The kernel in Eq.~\ref{eq:linearized_gap} then becomes
\begin{align}
& \tilde{\Gamma}_{k,k'} \simeq \\
& \Gamma_{BB}^{\rm RPA} (0) \frac{ \varepsilon_{k,+} \varepsilon_{k',+} + [\Delta/2]^2 }{2\varepsilon_{k,+} \varepsilon_{k',+}} + \Gamma_{AB}^{\rm RPA} (0)  \frac{ \Re\left[t_{\rm F}^2 f^*(k') f(k)\right] }{2\varepsilon_{k,+} \varepsilon_{k',+}} \notag
\end{align}
which holds up to $q^2$ corrections. Then, expanding the functions of $k$ and $k'$ near the $\pm K$ points, we find 
\begin{equation}
\tilde{\Gamma}_{k,k'} \simeq \Gamma_{BB}^{\rm RPA} (0) + 0 , 
\end{equation}
which also holds up to quadratic corrections of the form $|k \pm \vec{K}|^2$ or $|k' \pm \vec{K}|^2$. This gives the effective attractive interaction $\tilde{\Gamma}_{k,k'} \simeq -g_0$ between the two valleys. Finally, we need to expand the dispersion relation near the $\pm K$ points to find the effective mass
\begin{equation}
\varepsilon_{K+k,+} \simeq \frac{\Delta}{2} + \frac{|k|^2}{2m} , \quad m = \Delta_{\rm H} \big/ \left[ 3(at_{\rm F})^2 \right] ,
\end{equation}
which makes the connection with Eq.~\ref{eq:AppGapNumberEquation} complete. Note that we focus on intervalley pairing as intravalley in $s$-wave is forbidde for our spinless model ($N_f=1$).

\section{Evaluation of KL diagrams} \label{app:KLdiagrams}

The two-body vertex corrections $\delta \Gamma$ obtained from many-body perturbation theory are described by the diagrams in Eq.~\ref{eq:FiveKLDiagrams}, where we have used curvy lines for interaction events 
\begin{equation}
\begin{tikzpicture}[baseline=0cm,scale=0.8]
\draw (-0.5,0.5) -- (0.5,0.5); \draw (-0.5,-0.5) -- (0.5,-0.5); \draw[snake it] (0.,0.5) -- (0.,-0.5); 
\node at (-1.3,0.7) {\scriptsize $(k_a,s_a)=a$}; \node at (-1.3,-0.7) {\scriptsize $(k_b,s_b)=b$}; \node at (1.3,-0.7) {\scriptsize $c=(k_c,s_c)$}; \node at (1.3,0.7) {\scriptsize $d=(k_d,s_d)$}; \node at (1.15,0.0) {\scriptsize $q = k_d-k_a$}; 
\end{tikzpicture} \equiv \Gamma_{dc,ba}  , 
\end{equation}
and straight lines for the bare HF propagators
\begin{equation} 
\begin{tikzpicture}[baseline=0cm,scale=0.8]
\draw (-0.5,0.) -- (0.5,0.); \node at (0,0.2) {\scriptsize $a,b$};
\end{tikzpicture} \equiv G_{b,a}^{(0)} (i \omega_m) = \delta_{k_a,k_b} \sum_{n=\pm} \frac{\Psi_{k_b,n}^{s_b \, *}\Psi_{k_a,n}^{s_a}}{\xi_{k_a,n} - i \omega_m}  ,
\end{equation}
with $\omega_m$ a Matsubara frequency, and where $\xi_{k_a,n} = \varepsilon_{k_a,n} - \mu$ measures energies with respect to the chemical potential $\mu$. They are both expressed in the orbitals basis, leading to the important conservation $s_a=s_d$, $s_b=s_c$ in $\Gamma$, and to the single particle dressing $\Psi_{k_b,n}^{s_b \, *}\Psi_{k_a,n}^{s_a}$ in $G^{(0)}$. 

The five diagrams of Eq.~\ref{eq:FiveKLDiagrams} -- referred to as ladder (ldr), cross (crs), up and down wine glasses (uwg/dwg), and bubble (bbl) diagrams due to their form -- contribute to  the renormalization of the two-body scattering vertex~\cite{kohn1965new,maiti2013superconductivity,kagan2014kohn}:
\begin{subequations} 
\begin{equation}
\begin{tikzpicture}[baseline=0cm,scale=0.8]
\draw (-0.85,0.5) -- (0.85,0.5); \draw (-0.85,-0.5) -- (0.85,-0.5); 
\draw[snake it] (-0.5,0.5) -- (-0.5,-0.5); \draw[snake it] (0.5,0.5) -- (0.5,-0.5);
\node at (-0.75,0.75) {$1$}; \node at (-0.75,-0.75) {$2$}; \node at (0.75,-0.75) {$3$}; \node at (0.75,0.75) {$4$}; 
\node at (0.,0.75) {$d,a$}; \node at (0,-0.75) {$c,b$}; 
\end{tikzpicture} \; \delta \Gamma_{43,21}^{\rm ldr} = \frac{-1}{N} \sum_{abcd} \chi_{dc,ba}^- \Gamma_{43,ba} \Gamma_{dc,21} , 
\end{equation}
\begin{equation}
\begin{tikzpicture}[baseline=0cm,scale=0.8]
\draw (-0.85,0.5) -- (0.85,0.5); \draw (-0.85,-0.5) -- (0.85,-0.5); 
\draw[snake it] (-0.5,0.5) -- (0.5,-0.5); \draw[snake it] (0.5,0.5) -- (-0.5,-0.5);
\node at (-0.75,0.75) {$1$}; \node at (-0.75,-0.75) {$2$}; \node at (0.75,-0.75) {$3$}; \node at (0.75,0.75) {$4$}; 
\node at (0.,0.75) {$d,a$}; \node at (0,-0.75) {$c,b$}; 
\end{tikzpicture} \; \delta \Gamma_{43,21}^{\rm crs} = \frac{-1}{N} \sum_{abcd} \chi_{dc,ba}^+  \Gamma_{4c,2a} \Gamma_{d3,b1} , 
\end{equation}
\begin{equation}
\begin{tikzpicture}[baseline=-0.1cm,scale=0.8]
\draw (-0.85,0.5) -- (-0.5,0.5) -- (0,0) -- (0.5,0.5) -- (0.85,0.5); \draw (-0.85,-0.65) -- (0.85,-0.65); 
\draw[snake it] (0,-0.65) -- (0,0); \draw[snake it] (0.5,0.5) -- (-0.5,0.5);
\node at (-0.75,0.75) {$1$}; \node at (-0.75,-0.9) {$2$}; \node at (0.75,-0.9) {$3$}; \node at (0.75,0.75) {$4$}; 
\node[rotate=-45] at (-0.5,0.) {$d,a$}; \node[rotate=45] at (0.5,0.) {$c,b$}; 
\end{tikzpicture} \; \delta \Gamma_{43,21}^{\rm uwg} = \frac{-1}{N} \sum_{abcd} \chi_{dc,ba}^+  \Gamma_{c3,2a} \Gamma_{d4,b1} , 
\end{equation}
\begin{equation}
\begin{tikzpicture}[baseline=0cm,scale=0.8]
\draw (-0.85,-0.5) -- (-0.5,-0.5) -- (0,0) -- (0.5,-0.5) -- (0.85,-0.5); \draw (-0.85,0.65) -- (0.85,0.65); 
\draw[snake it] (0,0.65) -- (0,0); \draw[snake it] (0.5,-0.5) -- (-0.5,-0.5);
\node at (-0.75,0.9) {$1$}; \node at (-0.75,-0.75) {$2$}; \node at (0.75,-0.75) {$3$}; \node at (0.75,0.9) {$4$}; 
\node[rotate=45] at (-0.5,-0.1) {$d,a$}; \node[rotate=-45] at (0.5,-0.1) {$c,b$}; 
\end{tikzpicture} \; \delta \Gamma_{43,21}^{\rm dwg} = \frac{-1}{N} \sum_{abcd} \chi_{dc,ba}^+  \Gamma_{4c,a1} \Gamma_{d3,b2} , 
\end{equation}
\begin{equation} \label{eq:KLbubblecontributions}
\begin{tikzpicture}[baseline=0cm,scale=0.8]
\draw (0,0) circle (0.3); 
\draw (-0.5,-0.6) -- (0.5,-0.6); \draw (-0.5,0.6) -- (0.5,0.6);
\draw[snake it] (0,0.6) -- (0,0.3); \draw[snake it] (0,-0.6) -- (0,-0.3);
\node at (-0.4,0.85) {$1$}; \node at (-0.4,-0.85) {$2$}; \node at (0.4,-0.85) {$3$}; \node at (0.4,0.85) {$4$}; 
\node at (-0.43,-0.25) {$d$}; \node at (0.4,-0.25) {$b$}; \node at (-0.43,0.25) {$a$}; \node at (0.4,0.25) {$c$}; 
\end{tikzpicture} \;\; \delta \Gamma_{43,21}^{\rm bbl} = \frac{N_f}{N} \sum_{abcd} \chi_{dc,ba}^+  \Gamma_{4c,a1} \Gamma_{d3,2b} , 
\end{equation}
\end{subequations}
where we recall that the sums labeled by the generalized indices $abcd$ runs over all sublattice indices $s_a, s_b, s_c, s_d \in\{A,B\}$ and all momenta $k_a, k_b, k_c, k_d \in {\rm BZ}$. The particle-particle and particle-hole susceptibilities are computed by the Matsubara sums
\begin{equation}
\chi_{dc,ba}^{\epsilon = \pm} =  \frac{1}{\beta} \sum_{\omega_m} G_{d,a}^{(0)} (i \epsilon \omega_m) G_{c,b}^{(0)} (i \omega_m) ,
\end{equation}
which gives the result of Eq.~\ref{eq:SusceptibilitiesKL}. The resulting vertex corrections 
\begin{equation} 
\delta \Gamma = \delta \Gamma^{\rm ldr} + \delta \Gamma^{\rm crs} + \delta \Gamma^{\rm uwg} + \delta \Gamma^{\rm dwg} + \delta \Gamma^{\rm bbl} ,
\end{equation}
appear in Eq.~\ref{eq:AllKLTerms} of the main text.

\section{Diagrammatic description of the origin of an attractive interaction.} \label{app:diagrams}

\begin{figure}
\centering
\includegraphics[width=0.9\columnwidth]{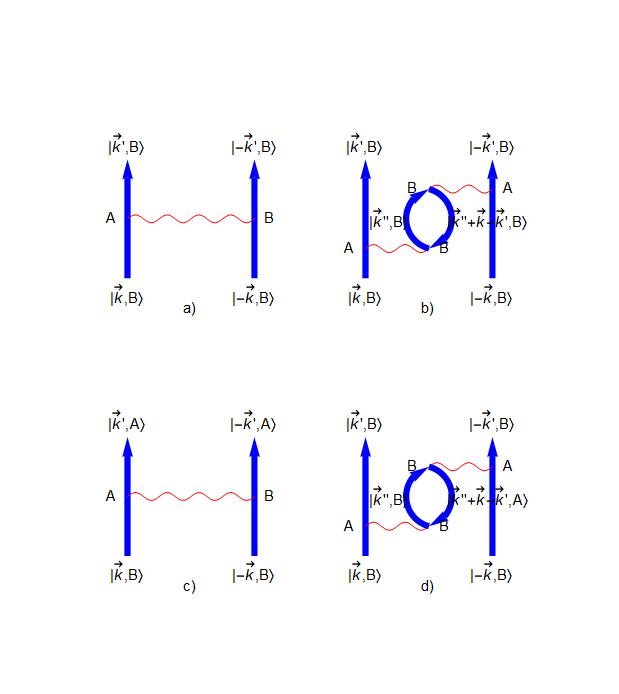}
\caption{Some diagrams which describe the scattering of Cooper pairs, and appear in the superconducting kernel in Eq.~\ref{eq:kernelnumerics}. The band labels $A$ and $B$ stand for the sublattice where the band is mostly localized. States $| k , A \rangle , |k' , A \rangle$ lie near the Fermi surface of one valley, and states $| - k , A \rangle , - |k' , A \rangle$ lie near the Fermi surface of the other valley. The states within the bubbles in diagrams b) and d) belong to the same valley, either $K$ or $K'$.  The overall momentum transfer is such that $| q | = | k - k' | \ll k_F$.
}
\label{fig:diagrams_RPA}
\end{figure}

In the following, we give a simplified description of the origin of attractive terms at small momentum transfer, $| q | \ll k_F$, in the superconducting kernel of the model Eq.~\ref{eq:kernelnumerics}. We assume, for simplicity, a small hole doping, which leads to two isotropic bands, one per valley, with the filling depermined by the Fermi wavevector, $k_F$. The partially occupied bands are mostly localized in sublattice $B$, and there are two empty bands at energies of order $\Delta$.

Some low order ladder and bubble diagrams which describe interaction induced scattering of Cooper pairs, and which contribute to the superconducting kernel in Eq.~\ref{eq:kernelnumerics} are shown in Fig.~\ref{fig:diagrams_RPA}. The bare interaction at small momenta is $\lim_{| q | \rightarrow 0} V( q ) = 3 V$, where the factor 3 arises from the number of nearest neighbors. The interaction is between electrons residing at $A$ and $B$ sites, so that the vertices involve form factors describing the amplitude of the wavefunctions of the $A$ and $B$ bands at the $A$ and $B$ sites. The amplitude of a wavefunction of the $A$ band at the $A$ sublattice is $\approx 1$, and the amplitude on the $B$ sublattice is, on the average, of order $ t / \Delta$, and of order $ ( t / \Delta ) \times ( k_F a ) \sim ( t / \Delta ) \times \sqrt{n}$ for states near the points $K$ and $K'$ ( $k_F$ is the Fermi wavelength, $a$ is the lattice constant, and $n \sim ( k_F a )^2 \ll 1$ is the number of electrons per unit cell in the partially occupied band ). The reverse holds for wavefunctions in the $B$ band.

Diagram a) in Fig.~\ref{fig:diagrams_RPA} is due to the bare interaction. It is repulsive, of order $\sim 3 V \times ( t / \Delta )^2 \times n$. It has a multiplicity of 2, as the two interaction vortices can be exchanged. Diagram c) describes interband transitions, see~\cite{DL21}. It gives a contribution also of order  $\sim 3 V \times ( t / \Delta )^2 \times n$. It has a multiplicity of 2.

The bubble diagrams, b) and d), describe the screening of the bare interactions. Both are second order processes in perturbation theory, and lead to an attractive interaction. Diagram b) describes the contribution of the polarizability of the partially occupied valence band, modulated by the weight of the the wavefunctions of this band on the $B$ sublattice. Its contribution is $\sim - ( 3 V )^2 \times ( t / \Delta )^4 \times ( \Delta / t^2 )$, where the factor $\Delta / t^2$ describes approximately the polarization of the valence band for $| q | \rightarrow 0$. The contribution of diagram d) is $\sim - (3 V )^2 \times ( t / \Delta )^2 \times ( 1 / \Delta)$, where the factor $1 / \Delta$ describes the energy cost of making a transition to the conduction band. The multiplicity of these diagrams is the number of electron flavors times 4, from the all possible exchanges of vertices.

The analysis of the diagrams in Fig.~\ref{fig:diagrams_RPA} suggests that the superconducting kernel, for $| q | \ll k_F$, becomes attractive for $3 V \sim \Delta$, in agreement with the results in the main text. For $n \ll 1$, the leading diagram is b), leading to a pairing interaction of order $( V^2 t^2 ) / \Delta^3 $, as in Eq.~\ref{eq:g0}.

\section{Hartree and Fock corrections} \label{app:HartreeFock}

The Hartree and Fock diagrams are graphically represented as
\begin{equation} \label{eq:HartreeFockDiagrams}
\begin{tikzpicture}[baseline=0cm,scale=0.8]
\draw (0,0.75) circle (0.25); \draw (-0.5,0.) -- (0.5,0.); \draw[snake it] (0.,0.) -- (0.,0.5);
\end{tikzpicture} , \quad 
\begin{tikzpicture}[baseline=0cm,scale=0.8]
\draw (-1,0.) -- (1,0.); \draw[snake it] (-0.5,0.) to [bend left=45] (0.,0.5) to [bend left=45] (0.5,0.);
\end{tikzpicture} .
\end{equation}
Up to a global shift of chemical potential, the Hartree contribution reads
\begin{equation}
h^{\rm H} ({q}) = \frac{3V}{2} \begin{bmatrix} -\delta & 0 \\ 0 & \delta
\end{bmatrix} , 
\end{equation}
where $\delta = (N N_f)^{-1} \langle \sum_{r\in A} n_r -\sum_{r\in B} n_r \rangle$ is the sublattice polarization. 
The Hartree contribution can thus be simply taken into account through a redefinition of the sublattice potential difference $\Delta_{\rm H} = \Delta + 3V \delta$. 
The Fock term is purely off-diagonal and assumes the form
\begin{equation}
h^{\rm F} ({q}) = - V \sum_{j=1}^3 \begin{bmatrix} 0 & e^{i({q}\cdot{u}_j)} t_j \\ e^{-i({q}\cdot{u}_j)} t_j^* & 0 \end{bmatrix} , 
\end{equation}
with $t_j = (N N_f^2)^{-1} \sum_{{r} \in B} \langle c_{{r}}^\dagger c_{{r}+{u}_j} \rangle$. 
As long as the system does not spontaneously break the $C_3$ symmetry of the original model, the Fock correction can be accounted for by a redefinition of the NN tunneling amplitude $t_{\rm F} = t + V t_0$, with $t_0 = (t_1+t_2+t_3)/3$. 

Note that these contributions could have equivalently derived by performing the standard Hartree-Fock substitution 
\begin{align}
n_{r,\sigma} n_{r',\sigma'} & = \langle n_{r',\sigma'} \rangle n_{i,\sigma} + \langle n_{r,\sigma} \rangle n_{j,\sigma'} \\ & - \langle c_{r,\sigma}^\dagger c_{r',\sigma'} \rangle c_{r',\sigma'}^\dagger c_{r,\sigma} - \langle c_{r',\sigma'}^\dagger c_{r,\sigma} \rangle c_{r,\sigma}^\dagger c_{r',\sigma'} \notag \\
& - \langle n_{r,\sigma} \rangle  \langle n_{r',\sigma'} \rangle + | \langle c_{r,\sigma}^\dagger c_{r',\sigma'} \rangle |^2  , \notag
\end{align}
in the interaction term of Eq.~\ref{eq:NfFlavorModel}, and then discarding constant terms. 

In conclusion, the Hartree-Fock corrections to the single particle Hamiltonian can be simply accounted for by a redefinition of the sublattice potential difference $\Delta \to \Delta_{\rm H}$ and of the NN tunneling amplitude $t \to t_{\rm F}$. These new parameters should be self-consistently computed from 
\begin{subequations} \label{eq:selfconsistentHFequation} \begin{eqnarray}
\Delta_{\rm H} &=& \Delta + \frac{3V \langle  \sum_{r\in A} n_r - \sum_{r\in B} n_r \rangle}{N N_f}  , \\
t_{\rm F} &=& t + \frac{V \sum_{j=1}^3 \langle  \sum_{r\in B} c_r^\dagger c_{r+u_j} \rangle}{3 N N_f^2}  .
\end{eqnarray} \end{subequations}
The renormalized band dispersion and Bloch vectors follow from performing this substitution in Eqs.~\ref{eq:SingleParticleEigenEnergies} and~\ref{eq:SingleParticleBloch}.

In the weakly doped regime $x \ll 1$, the conduction band is almost empty, such that the expectation values in Eq.~\ref{eq:selfconsistentHFequation} are dominated by valence band contributions. This simplifies the self-consistent equations as
\begin{equation} \label{eq:selfconsistentHFequationunitfilling}
\frac{\Delta_{\rm H} - \Delta}{\Delta_{\rm H}} = \frac{3V}{2N} \sum_{{q}} \frac{1}{\varepsilon_{{q},+}}  , \quad
\frac{t_{\rm F} - t}{t_{\rm F}} = \frac{1}{6 N N_f} \sum_{{q}}
\frac{|f({q})|^2}{\varepsilon_{{q},+}}  .
\end{equation}
They allow to obtain approximate solutions for the Hartree and Fock corrections in the perturbative regimes studied in the main text. For instance, we observe that $|t_{\rm F}-t|\to 0$ when $N_f \to \infty$, as expected by inspection of the diagrams of Eq.~\ref{eq:HartreeFockDiagrams}. Indeed, the Hartree bubble can hold any of the $N_f$ fermionic flavors, while the flavor of the intermediate line in the Fock diagram is fixed. The other interesting limit is $t\ll\Delta$ used for the KE. In that regime, we find that electrons of the $n=1$ insulating state mostly live on $A$ sites, as expected from Sec.~\ref{sec:HybridizationExpansion} and described by $\Delta_{\rm H} = \Delta + 3V$ and $t_{\rm F} = t [1+V/(N_f \Delta_{\rm H})]$.

\section{Oversimplifying approximation} \label{app:FlatMetric}

We now detail the "flat-metric condition" introduced in the main text, and show that it leads to purely repulsive interaction, both in the spinless model Eq.~\ref{eq:OriginalSpinlessModel} and int he innteraction model for TBG.

In more detail, we assume that the form factors appearing in the definition of the susceptibility are diagonal in the band basis and do not vary over the BZ, as they would if correlated hopping terms were present. For the spinless model Eq.~\ref{eq:OriginalSpinlessModel} and twisted bilayer graphene, this respectively corresponds to
\begin{align} 
\label{approx_mat_2}
\Psi_{k,m}^{i \, *} \Psi_{k,n}^i & \approx \delta_{n,m} \overline{\Psi_{k,n}^{i \, *} \Psi_{k,n}}\big|_{k  \in {\rm BZ}} = \delta_{n,m} f_n^i , \\
\braOket{k,m}{ e^{iG r} }{k,n} & \approx \delta_{n,m} \overline{ \braOket{k , n}{e^{iG r} }{ k , n}} \big|_{k  \in {\rm BZ}} =  \delta_{n,m} f_n^G . \notag
\end{align}
Focusing on the conduction band, the static susceptibility coefficients can be written as
\begin{equation}
\chi_{\alpha,\beta}^c (q) = \bar{\chi} (q) f_{c}^\alpha f_{c}^\beta , \quad \bar{\chi} (q) = \!\!\! \sum_{k \in {\rm BZ}} \!\! \frac{f_\beta(\xi_{k + q , n}) - f_\beta (\xi_{k , n})}{\xi_{k + q , n} - \xi_{k , n}} ,
\label{approx_susc}
\end{equation}
with $c$ refers to the conduction band, and where, depending on the model, $\alpha$ and $\beta$ either denote orbital indices ($A/B$) or moiré reciprocal lattice vectors ($G$). 

In the spinless model of Eq.~\ref{eq:OriginalSpinlessModel}, this approximation leads to a screened RPA vertex 
\begin{widetext}
\begin{equation}
\Gamma^{\rm RPA} (q) \approx \frac{1}{1-2V\Re[f(q)] f_c^A f_c^B \bar{\chi}(q)} \begin{bmatrix}
|V f (q) f_c^B|^2 \bar{\chi} (q) & V f(q) [1-V f^*(q) f_c^A f_c^B \bar{\chi}(q)] \\ V f^*(q) [1-V f(q) f_c^A f_c^B \bar{\chi}(q)] & |V f(q) f_c^A|^2 \bar{\chi} \end{bmatrix} . 
\end{equation}
\end{widetext}
In turn, this vertex produces an effective Cooper interaction (defined in Eq.~\ref{eq:linearized_gap})
\begin{equation}
V_{c,k'}^{c,k}  \approx \frac{2 V \Re[f(k'-k)] f_c^A f_c^B}{1 - 2 \bar{\chi} (k'-k) V \Re[f(k'-k)] f_c^A f_c^B} , 
\end{equation}
which is positive, \textit{i.e.} repulsive, for small momentum difference since $\bar{\chi}(q\to 0)<0$, $f(q=0)=3$ and $f_c^A, f_c^B > 0$. As a result, washing out the interband contributions and the correlated hopping terms is too restrictive to explain the results obtained throughout this paper.

Similarly, for twisted bilayer graphene, following similar steps, we find that the approximation of Eq.~\ref{approx_mat_2} leads to the 
\begin{equation}
\Gamma_{G , G'}^{\rm RPA} ( q ) \approx V^0_{G } ( q ) \delta_{G , G'} + \frac{V^0_{G} ( q )  V^0_{G'} ( q ) \bar{\chi} ( q ) f_{G} f_{G'}}{1 - \bar{\chi} ( q )  \sum_{G''} V^0_{G''} ( q ) f_{G''}^2} , 
\end{equation}
where $V^0$ denotes the bare Coulomb repulsion, and $\bar{\chi} ( q )$ is given by an approximation similar to that in Eq.~\ref{approx_susc}. Projecting the latter in the conduction band, we find an effective Cooper potential  
\begin{equation}
V_{c,k'}^{c,k} \approx g^2(k,k') \frac{\sum_{G} V^0_{G} ( k'-k ) f_{G}^2}{1-\bar{\chi} (k'-k) \sum_{G} V^0_{G} ( k'-k ) f_{G}^2 } , 
\end{equation}
where $g(k,k')$ is a band structure dependent coefficient (see Refs.~\cite{Ceae2107874118} for more details). Because $\tilde{\chi} ( q ) < 0$, all the matrix elements of this potential are positive, describing a net repulsion.

\end{document}